\pdfoutput=1
\documentclass{aa}  

\usepackage{graphicx}
\usepackage{txfonts}
\usepackage{natbib}
\bibpunct{(}{)}{;}{a}{}{,} 

\begin{document} 

\title{iNNterpol\thanks{\texttt{https://github.com/cwestend/iNNterpol}}: High-precision interpolation of stellar atmospheres with a deep neural network using a 1D convolutional auto encoder for feature extraction}

\subtitle{}

\author{C. Westendorp Plaza
          \inst{1}\inst{2}\fnmsep\thanks{carlos.westendorp@iac.es}
          \and
          A. Asensio Ramos\inst{1}\inst{2}
          \and
          C. Allende Prieto\inst{1}\inst{2}
          }

\institute{Instituto de Astrof\' isica de Canarias, C/V\'{\i}a L\'actea s/n, E-38205 La Laguna, Tenerife, Spain\\
              \email{carlos.westendorp@iac.es} \email{andres.asensio@iac.es}  \email{carlos.allende.prieto@iac.es}
         \and
             Departamento de Astrof\' isica, Universidad de La Laguna, E-38206 La Laguna, Tenerife, Spain\\
             }

\date{Received march 2023; accepted may 2023}

 
 \abstract
   {Given the widespread availability of grids of models for stellar atmospheres, it is necessary to recover intermediate atmospheric models by means of accurate techniques that go beyond simple linear interpolation and capture the intricacies of the data.}
   {Our goal is to establish a reliable, precise, lightweight, and fast method for recovering stellar model atmospheres, that is to say the stratification of mass column, temperature, gas pressure, and electronic density with optical depth given any combination of the defining atmospheric specific parameters: metallicity, effective temperature, and surface gravity, as well as the abundances of other key chemical elements.}
   {We employed a fully connected deep neural network which in turn uses a 1D convolutional auto-encoder to extract the nonlinearities of a grid using the ATLAS9 and MARCS model atmospheres.}
   {This new method we call iNNterpol effectively takes into account the nonlinearities in the relationships of the data as opposed to traditional machine-learning methods, such as the light gradient boosting method (LightGBM), that are repeatedly used for their speed in well-known competitions with reduced datasets. We show a higher precision with a convolutional auto-encoder than using principal component analysis as a feature extractor. We believe it constitutes a useful tool for generating fast and precise stellar model atmospheres, mitigating convergence issues, as well as a framework for future developments.
   The code and data for both training and direct interpolation are available online for full reproducibility and to serve as a practical starting point for other continuous 1D data in the field and elsewhere.}
  {}
 
 \keywords{Methods: numerical --
                Methods: data analysis --
                Stars: atmospheres --
                Catalogs
               }
 \titlerunning{iNNterpol: High-precision interpolation of stellar atmospheres with a deep neural network}

 \maketitle
%

\section{Introduction}

In order to study stellar spectra from observed data, a widespread approach is to resort to theoretical model atmospheres. They represent tabulated thermodynamical quantities such as density, temperature, pressure, electron number density, and opacity as a function of optical depth for a wide range of stellar atmospheric parameters, such as effective temperature, surface gravity, and chemical composition. Commonly, a grid of such models are painstakingly calculated for a large set of parameters \citep{ATLAS9_Kirby2011}. 
Intermediate values of these parameters are obtained by interpolating in this grid. Since these grids span a wide range of stellar atmospheres, where different physical processes have to be taken into account, the question arises of whether we can go beyond a simple linear interpolation among these models. 

The purpose of this work is to investigate a way to recover all the possible atmospheres within a specific grid in a fast way and also with great precision, going beyond a straightforward linear interpolation. The method is presented here, where the nonlinear relations within the data were recovered by means of a neural network (NN). 

We applied these techniques to two well-known families of models that effectively cover most of the parameter space of effective temperatures, surface gravities, and metallicity ratios such as the ATLAS9 and MARCS collections of model atmospheres. The ATLAS9 family consists of over 853,000 models which are obtained by the well-known code of \cite{ATLAS_Kurucz1979} enhanced by new opacities and covering a wider range of stellar compositions. The 
MARCS code (\citealt{MARCS_2008}) with an augmented metallicity subgrid provides over 380,000 models. Both grids of models are described in \cite{Atlas9_Meszaros2012}.

These series of models serve as a starting point for our technique, which by no means is restricted to them. We firmly believe that this type of NN interpolator can be very useful for any other grid of models when the parameter space is sufficiently covered, as we discuss below.

To provide a more comprehensive set of models and access the latest updates, we extended our technique to include the PHOENIX model atmospheres (\citealt{PHOENIX_2013}). However, compared to the other two models, PHOENIX has a much smaller parameter space as it does not include the carbon abundance as a specific parameter. This leads to a significantly smaller number of available models, which is roughly ten times less than in the other two families.

Neural networks, which have shown great success in various fields from image recognition and generation to natural language processing, have also been used in different aspects of astronomy (e.g., see \citealp{IACWS_2019arXiv_Baron}). In our particular case, we wanted to apply a NN to a series of smooth stellar atmospheric models, in a way that the NN could learn the characteristic features of these models for the initial stellar parameters (effective temperature, surface gravity, and metal abundances). Given new values of these parameters, the NN would be able to generate an atmosphere based on the features learned with sufficient precision in order to be used in subsequent tasks such as calculating stellar synthetic spectra. In order to reduce the number of parameters and effectively avoid the so-called curse of dimensionality, we needed to capture in some way the essence of these models with an effective feature-extraction method. In our case, even with a high number of models as we have (see Section~\ref{sec:data}), the great number of parameters (or dimensions) could render our attempts to obtain high precision futile. Additionally, the characteristic smooth nature of our data (all physical parameters) in optical depth implies that treating each value as independent to the next one is prone to imprecisions and sharp discontinuities.  

Initially we resorted to using principal component analysis (PCA) for dimensional reduction, as it is a well-known technique for this means and is described in Section~\ref{sec:pca}. Although PCA is known to be precise, especially with observed data (i.e., with noise), it is of linear nature. Applying singular value decomposition (SVD) to the matrix of stacked models and keeping a subset of the principal components is what we passed to our NN and this enabled us to train it. With our data being naturally smooth and not presenting any observational noise, there is no evident limit to the number of components to discard.

Taking advantage of recent progress in deep learning,  
we find that a convolutional auto-encoder (CAE) is particularly useful in this regard, as it can improve the results obtained with PCA even with the same number of components. We then proceed to elaborate the optimal solution or state of the art (SOTA), we call iNNterpol, described in Section~\ref{sec:cae}, that highly improves on PCA and gives a higher precision when tested with data not seen by the NN.

As a sidenote, we study the use of traditional machine-learning (ML) techniques, such as gradient boosting (GB, in particular LightGBM, \citealt{LightGBM_2017}), that in principle are optimal for tabular-like data. These techniques are simpler, faster, and have allowed for well-known ML challenges to be overcome (e.g., \citealt{lgbm_comparison2006}), albeit with a short time for development and thus employing limited data samples. We shall prove in Section~\ref{sec:gbm} that they are not entirely applicable for the problem at hand.

The aim of this work is to deliver both a useful tool and a method that can be applied to all sets of data of a 1D continuous nature, which are numerous in the natural sciences in general and particularly abundant in astrophysics. As the details here are really important and we know it is usually very hard to reproduce models in the literature from general descriptions, we provide all products, that is the tool with the full source code, together with the original data. These are available for full reproducibility of the present work and to serve as a starting point for future work\footnote{\texttt{https://github.com/cwestend/iNNterpol}}. 

This work is structured as follows: in Section~\ref{sec:data} we present the main characteristics of the model atmospheres involved, and we then explain how we we applied PCA for dimensionality reduction and used its results embedded in a NN as described in Section~\ref{sec:pca}. We first chose the ATLAS9 family of models as being the most extense. We then study the effect of a classical ML technique as GB in Section~\ref{sec:gbm} and finally present the optimal model we call iNNterpol, which we present in Section~\ref{sec:cae} by including a CAE as the feature extractor of our NN. Once we had the optimal solution for the ATLAS9 grid, we then applied it to the MARCS and PHOENIX grid of models, where small variations had to be performed. The difference for MARCS and PHOENIX with the ATLAS9 results are presented in Section~\ref{sec:3models}. The differences with classical linear interpolation are detailed in Section~\ref{sec:linear}.

\section{Description of the models}

\label{sec:data} 

We chose two families of models obtained through the ATLAS9 and MARCS codes, both described in detain in \citet{Atlas9_Meszaros2012}, as they are the most complete, homogeneous, and dense ones available in the literature. The ATLAS9 grid has solar-scaled metallicities [M/H] from -5 to 1.5, carbon [C/M] abundances, and $\alpha$-element [$\alpha$/M] variations ranging from -1.5 to 1. The effective temperatures span values ranging from 3,500 K to 30,000 K and log g from 0 to 5. These are 1D plane-parallel model atmospheres computed under the assumption of local thermodynamical equilibrium. For the MARCS grid of models, the metallicities [M/H] cover from -2.5 to 1.0, carbon [C/M] abundances, and $\alpha$-element [$\alpha$/M] variations ranging from -1.0 to 1.0. In the MARCS grid, the effective temperatures range from 2,500 K to 8,000 K and log g from -0.5 to 5.0.

For each combination of these five initial parameters, we have the stratification with optical depth of the mass column, temperature, gas pressure, and electron number density in 71 points for the ATLAS9 (56 points for MARCS models). 
We used the Rosseland optical depth scale:
\begin{equation}
\tau_{\rm Ross} (z) = - \int_{\inf}^{z}  \kappa_{\rm Ross} dz,
\end{equation}
where $ \kappa_{\rm Ross}$ is the Rosseland mean opacity
\begin{equation}
\kappa_{\rm Ross}^{-1} = \frac{\int_0^{\inf} \kappa_{\nu}^{-1} u(\nu, T) d\nu}{\int_0^{\inf} \ u(\nu, T) d\nu}
\end{equation}
with $u(\nu, T) = \frac{dB}{dT}$, the derivative of the Plank function, and $\kappa{\nu}$ is the frequency-dependent opacity.

Not all combinations are physically viable or yield stable stellar atmospheres. For instance, at high effective temperatures, there are only models with high gravities. We have roughly over 853,000 models for ATLAS9 and 381,000 for MARCS covering most stellar types so they sample the parameter space sufficiently.

We also tried to apply a similar technique to the PHOENIX grid, but the results are not as precise as for the other two, as we discuss below. We think this is due to the fact that these models lack an order of magnitude as they do not sample the carbon abundances, thus providing a total of only 47,000 models. For the PHOENIX grid, the range of metallicities [M/H] varies from -4.0 to 1.0, $\alpha$-element [$\alpha$/M] variations range from -0.4 to 1.2, the effective temperatures range from 2,300 K to 15,000 K, and log g varies from -0.5 to 6.5.

\section{Methods and results}

We applied the following techniques on the ATLAS9 family of models as they are the largest one and they cover more parameter space. An adaptation of the resulting model applied to the other two datasets is discussed in Sect. \ref{sec:3models}.

\subsection{PCA in a NN}
\label{sec:pca} 

Principal component analysis, or Karhunen-Lo\`eve transformation, is a multivariate statistical technique that has been widely used in stellar (see \citealt{PCA_JMunyoz_2013, PCA_Carroll2007}) and solar spectroscopy (\citealp{PCA_Marian2008, Asensio_Ramos_2007}) for quite some time \citep{PCA_Rees2000, PCA_Stellar_BailerJones1998}. It is a very fast and efficient way of reducing the feature space by means of a linear transformation of the data that finds the direction along which the variance is maximum. Removing second-order dependencies yields an orthonormal basis for which the directions are uncorrelated. It effectively reduces the dimensionality of the data as we chose to keep a subset of these components that is able of reconstructing the original data within a certain error. PCA works best when the dimensions in the original data space are related to each other and when nonlinear effects are small, but it is just an assumption since this is basically unknown a priori. 

In the case of noisy data as in real spectra, it is natural to decide on the number of components or eigenvalues to keep since the rest should ideally contain only the information about the noise. In our case, with smooth and noiseless synthetic models, the criterion chosen was to retain as many components as needed to recover the original data within a specific error (ideally less than 2\% RMS).

\begin{figure}
\resizebox{\hsize}{!}{\includegraphics{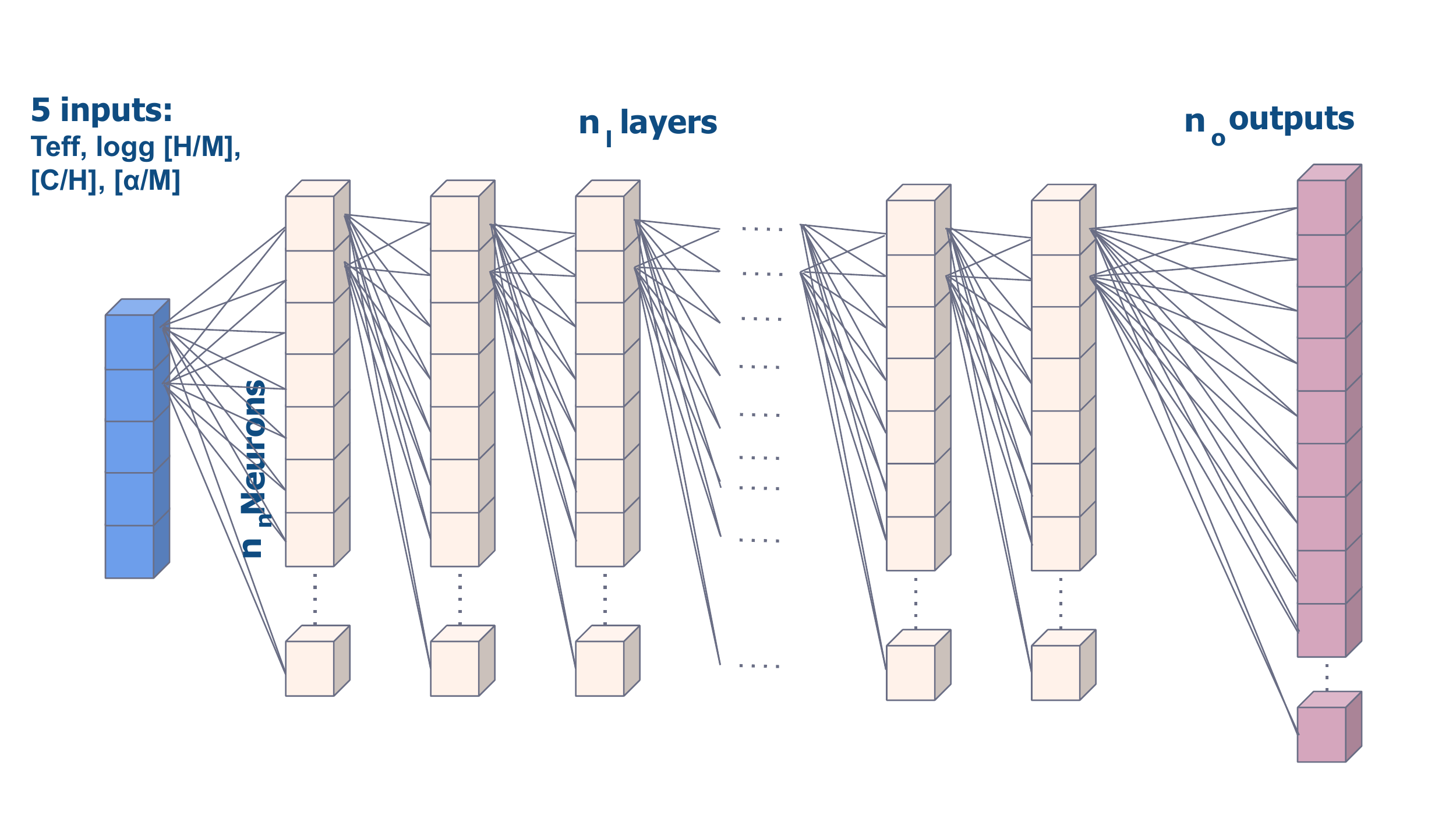}}
\caption{General architecture of our fully connected NN model (some connections have been omitted for clarity). The input parameters are the five values of the grid, effective temperatures Teff, surface gravity log g, and metalicities [M/H], [C/M], and [$\alpha$/M]. The n$_o$ output values are the result of the dimensionality reduction used: for PCA the 12 components of each parameter for a specific model atmosphere (48 in total) and when using the encoder of our CAE 48, 64, and up to 71 which constitutes the best configuration. The values of n$_n$ neurons per layer and n$_l$ layers are described in Table~\ref{tab:NN}.}
\label{fig:NN_gslide}
\end{figure}

For our model atmospheres and especially due to the smooth nature of our data, PCA not only provides a dimensional reduction, which is a natural starting point for a NN analysis \citep{bishop1995}, but it also ensures that the recovered stratification of all parameters should be both smooth and continuous. To this end we set up our NN as described in Fig.~\ref{fig:NN_gslide}, where the input are our five model grid parameters (effective temperature, surface gravity, and metal abundances: total [M/H], carbon abundance [C/M], and $\alpha$ elements [$\alpha$/M]) and the output are the first 12 PCA components for each of the physical parameters on which SVD was applied. We worked with the logarithm of these quantities to keep the variations within a similar range for all quantities and for the NN to be able to train equivalent weights. We resorted to the highest numerical precision available provided by the language (in our case numpy.longdouble, or 128-bit extendedprecision, i.e., quadruple precision or 16-byte real numbers) in applying SVD to minimize the rounding error when recovering the original quantities. We decided 12 components were precise enough since the error recovering the parameters was within the  2\% RMS error and previous calculations with only nine PCA components for each of the four physical quantities (PCA36-NN) gave a precision which was only slightly worse but which overestimated the predicted values for the temperature stratification in higher layers (see Fig.~\ref{fig:DTemp_chk36_M0_mp05_lgg45} in Appendix~\ref{sec:extra}). The rest of the hyperparameters in the NN were estimated by trial and error (number of layers, number of neurons per layer, activation functions, learning rates, epoch number, and batch size). The optimal values are shown in Table~\ref{tab:NN}. The parameter inputs for each model were initially normalized by the absolute maximum value in their range. We also experimented with values in the batch size, and found that better convergence was obtained using a small size of 64, which in out network of few input parameters and larger output ones with many layers resulted in a slow training as the usage of the GPUs was not optimized. This meant that for 100 epochs, each training took around 3 hours of computing time on an NVidia Tesla P100 with 12GB, which is perhaps what also hinders the use of NNs as opposed to classical ML where it takes around 10-15 minutes as we shall see in Section~\ref{sec:gbm}.

We trained the NN using the usual strategy of employing 80$\%$ of the data for training, 10$\%$ for validating, and 10$\%$ as test data, which was all unseen by the NN to evaluate the quality of the predictions. All were chosen randomly to avoid training and validating with a specific portion of parameter space with could have certain peculiar characteristics.

\begin{table}
\centering
\caption{Range of parameters tested for the fully connected NN and optimal value found for PCA with 12 components and using a CAE with a bottleneck layer of 48, 64, and 71 (CAE48, CAE64, and the optimal CAE71 we call iNNterpol). This all pertains to ATLAS9 data; for more details on MARCS and PHOENIX, readers can refer to Section~\ref{sec:3models}. The activation functions referred to are rectified linear unit (relu), leaky rectified linear unit (lkrelu), and exponential linear unit (elu).}
    \label{tab:NN}
    \small 
    \begin{tabular}{cccc} 
                \hline
                n$_l$ Layers & n$_n$ Neurons/layer & Activation & Batch size\\
                \hline
                8 to 22 & 20 to 80 & relu/lkrelu/elu & 256 to 32\\
                \hline
            Optimal values\\
            \hline
                (PCA36,48-NN)\\
                12 & 40 & lkrelu & 64\\
                (CAE48-NN)\\
                16 & 48 & lkrelu & 64\\
                (CAE71-NN\\ or iNNterpol)\\
                16 & 71 & elu & 64 \\
                \hline
    \end{tabular}
\end{table}

\begin{figure}
        \includegraphics[width=\columnwidth]{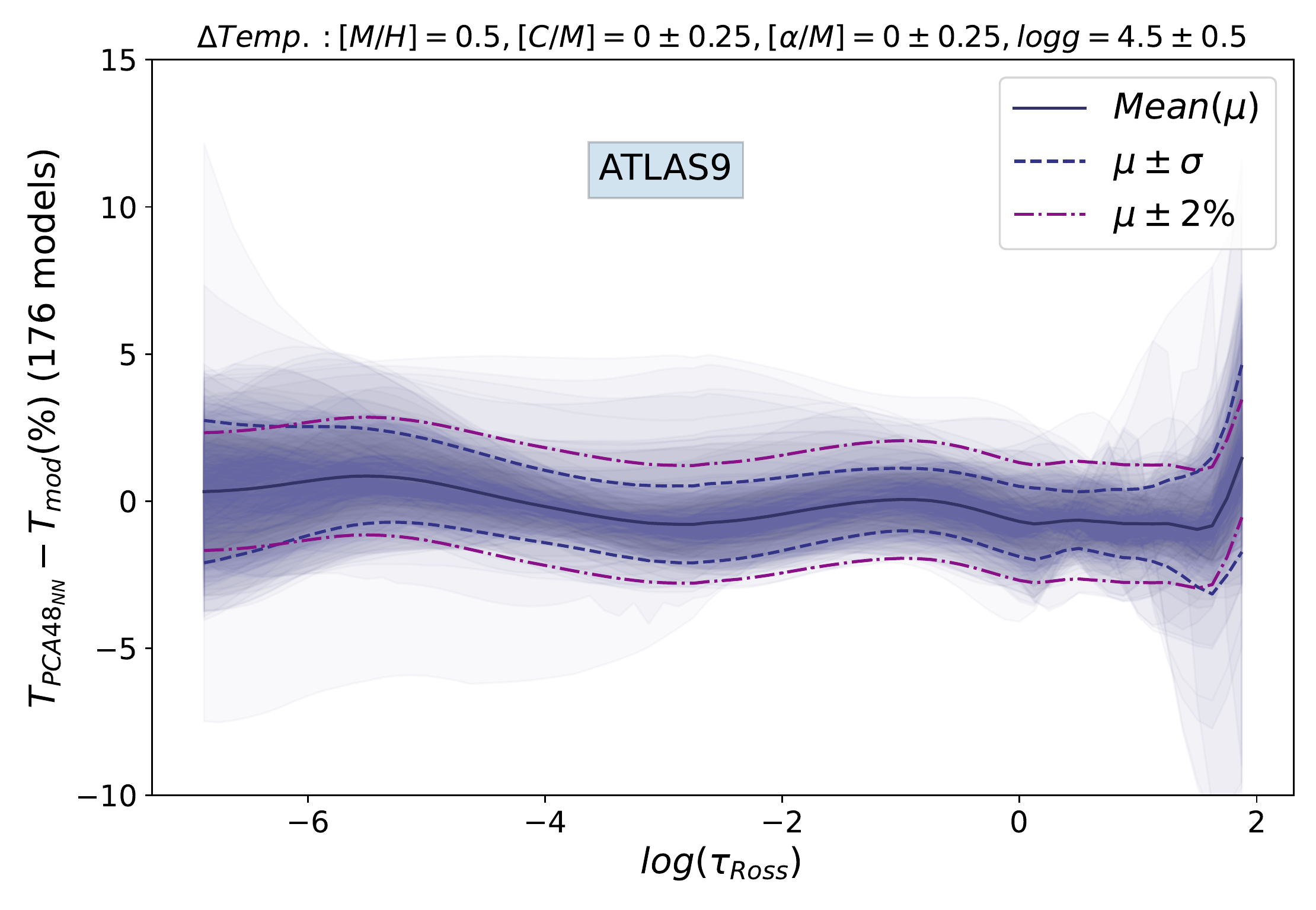}
    \caption{PCA48-NN (12 component PCA) differences in temperature as a percentage between the predicted models and the actual ones for points in the grid never seen by the NN for ATLAS9 model data. The full range of effective temperatures are covered (3,500 K to 30,000 K).}
    \label{fig:DTemp_chk48_M0_mp05_lgg45}
\end{figure}


The quality of the results of this PCA-NN using 12 PCA components for each physical quantity (PCA48-NN) is illustrated in Fig.~\ref{fig:DTemp_chk48_M0_mp05_lgg45} where the differences between the predicted values of the temperature stratification and the  "ground truth" (the actual model atmospheres) are shown for combinations of the input parameters in the test set. For these atmospheres of metal-rich dwarf stars (with [M/H] = 0.5, [C/M], [$\alpha$/M] = 0 $\pm$ 0.25, and log g = 4.5 $\pm$ 0.5), the errors lie well below the 2$\%$ value throughout the atmosphere.

\subsection{Gradient boosting}
\label{sec:gbm} 

A common question that always arises when applying deep networks such as the NN described above is what would happen when applying classical ML techniques. For this, we resorted to the well-known GB methods which are so extremely fast and known to work well with tabular data. 
In our case we resorted to the LightGBM implementation \citep{LightGBM_2017} which is known to be well suited for large datasets as in our case, is less prone to overfitting, as it has a built-in early-stopping mechanism and especially because it is more accurate.  

The results, after carrying out a complete hyperparameter search over 500 boosting rounds consuming from 10 to 15 minutes on a 32-core Xeon CPU~$@$2GHz in parallel mode (it can also work on GPU, but this was unnecessary), are shown in Table~\ref{tab:lgbm}. 
We followed the same strategy of training on 80$\%$ of the data, 10$\%$ for validation, and 10$\%$ for testing.

\begin{table}
    \centering
    \caption{Optimal parameters after a hyperparameter search for the lighGBM regressor.}
    \label{tab:lgbm}
    \small 
    \begin{tabular}{lccr} 
                \hline
                boost type & min data in leaf & subsample & max depth \\
                \hline
                gbdt & 1 & 0.5 & 15 \\
                \hline
                num estimators & metric & boost rounds & early stopping \\
                \hline
                8.000 & rmse & 500 & 10 \\
        \end{tabular}
\end{table}

\begin{figure}
        \includegraphics[width=\columnwidth]{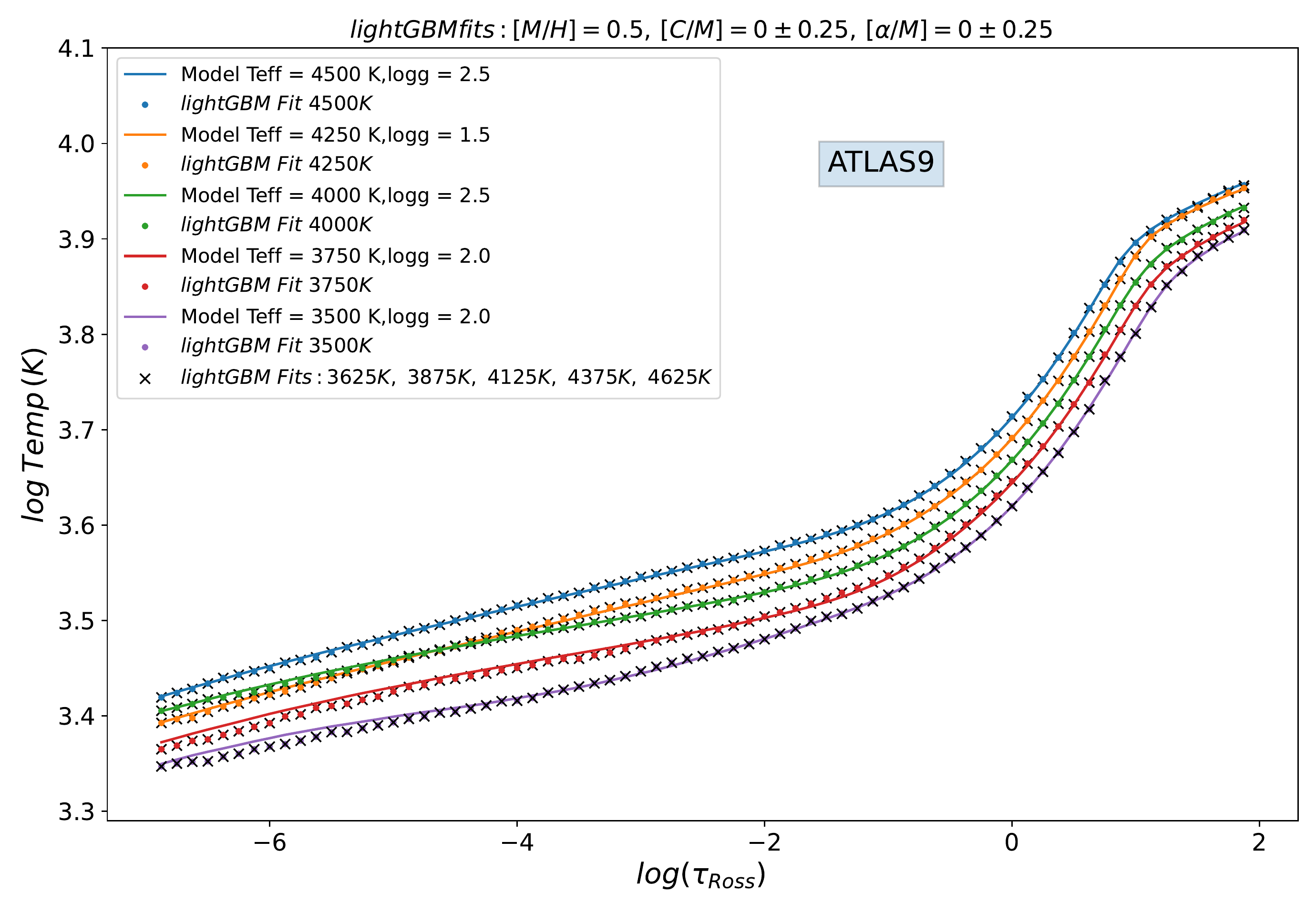}
    \caption{LightGBM fits to model data for test points. Solid lines correspond to actual model data and solid circles illustrate the fit of the LightGBM model to these parameters, predicting the temperature. The crosses show the predictions to the data just adding 125 K to the effective temperature parameter and leaving the rest of the paremeters unchanged. We note that the dots and crosses are predicted with the same values.}
    \label{fig:Temp_chklGBM_M0_3500_4675}
\end{figure}

The results are shown in Fig.~\ref{fig:Temp_chklGBM_M0_3500_4675} where different atmospheric models are plotted for different effective temperatures; all have solar-type abundances and low surface gravities. We can see that the predicted values of LightGBM are quite good and precise. The problem comes when predicting models for parameters that do not fall exactly on the grid nodes. This is what can be seen in Fig.~\ref{fig:Temp_chklGBM_M0_3500_4675} where we only vary the effective temperature adding 125 K, keeping the rest of the parameters (([M/H], [C/M], [$\alpha$/M], and log g)) with the same values as those of the model points. We note that the model values are equispaced in 250 K in effective temperature for this range. In this case, the predicted values do not fall between the two models with effective temperatures $\pm$ 125 K as would be expected (and as PCA-NN and CAE-NN interpolate), but exactly over the values of the model before. We conclude that at least in our case, for continuous models in a close equispaced parameter grid as we have, LightGBM acts only as a high-precision classifier (although it was set up to be a regressor). It somehow learns the structure of the grid of models and classifies the value of each temperature for each depth, but it is unable to learn and interpolate values in between. In this case, it works as a nearest neighbor interpolation so a 5D linear interpolation should be preferred as it can effectively recover intermediate values.

\subsection{Convolutional auto-encoder in the NN}
\label{sec:cae} 

To overcome the limitations of PCA regarding the possible nonlinear relations in the manifold that constitutes the parameter space, we resorted to a strategy that involves feature extraction based on a CAE. Here deep learning helps with the introduction of the auto-encoder (AE, \citealt{Autoencoder2006}). It has been shown that in the simplest case, an AE with {fully connected} layers and only linear activation on the output trains weights that span the same subspace as the one recovered by PCA \citep{Bourlard1988}, so in the worst case we would be obtaining the same gain as with PCA.

In a CAE the convolutional layer, as introduced by \cite{Fukushima1980}, contains units whose receptive fields cover a patch of the previous layer; the set of adaptive parameters or weight vector of such a unit is often called a filter. These convolutional layers can be stacked, each effectively retrieving features from the previous one. This technique was applied to handwritten images by \cite{CNN_LeCun89} and opened a broad field for image recognition that has since become the foundation of modern computer vision.  

Apart from the great success of convolutional neural networks (CNNs) which are now widely used in image processing and real-time recognition, the application to 1D data is paradoxically much less frequent. Besides notable examples in fields as diverse as medical sciences \citep{1DCNN_Med_Huang2018}, structural damage detection \citep{1DCNN_ABDELJABER2017}, or speech recognition \citep{1DCNN_Speech_Abdel2014}, many have yet to arise. Some even resort to converting the 1D signals to 2D structures (\citealt{1DCNN_Fault2016,1DCNN_Zihlmann2017}) to be able to apply the well-established techniques as used for images. The specificity of customizing the architecture of each CNN is time consuming and inherently challenging as there are no shortcuts. Additionally, the extensive hyperparameter space that needs to be sampled has surely limited the utilization of CNNs. An aspect of convolutions is that apart from reducing the size of the input data at each step in the filter dimension, they absorb the data size by increasing the number of channels, so there needs to be an additional way to create a bottleneck. In our case, we resorted to 1x1 convolution with a kernel size of 1 and stride of 1, made famous as used in the GoogleLeNet so-called inception architecture \citep{1x1Conv_GoogleLeNet2014}. This layer not only reduces the dimensions but also serves as a fully connected layer in the 1D CNN to which the activation functions enable it to learn even more characteristics of the data and enhance the power of the model.

\begin{figure}
        \includegraphics[width=\columnwidth]{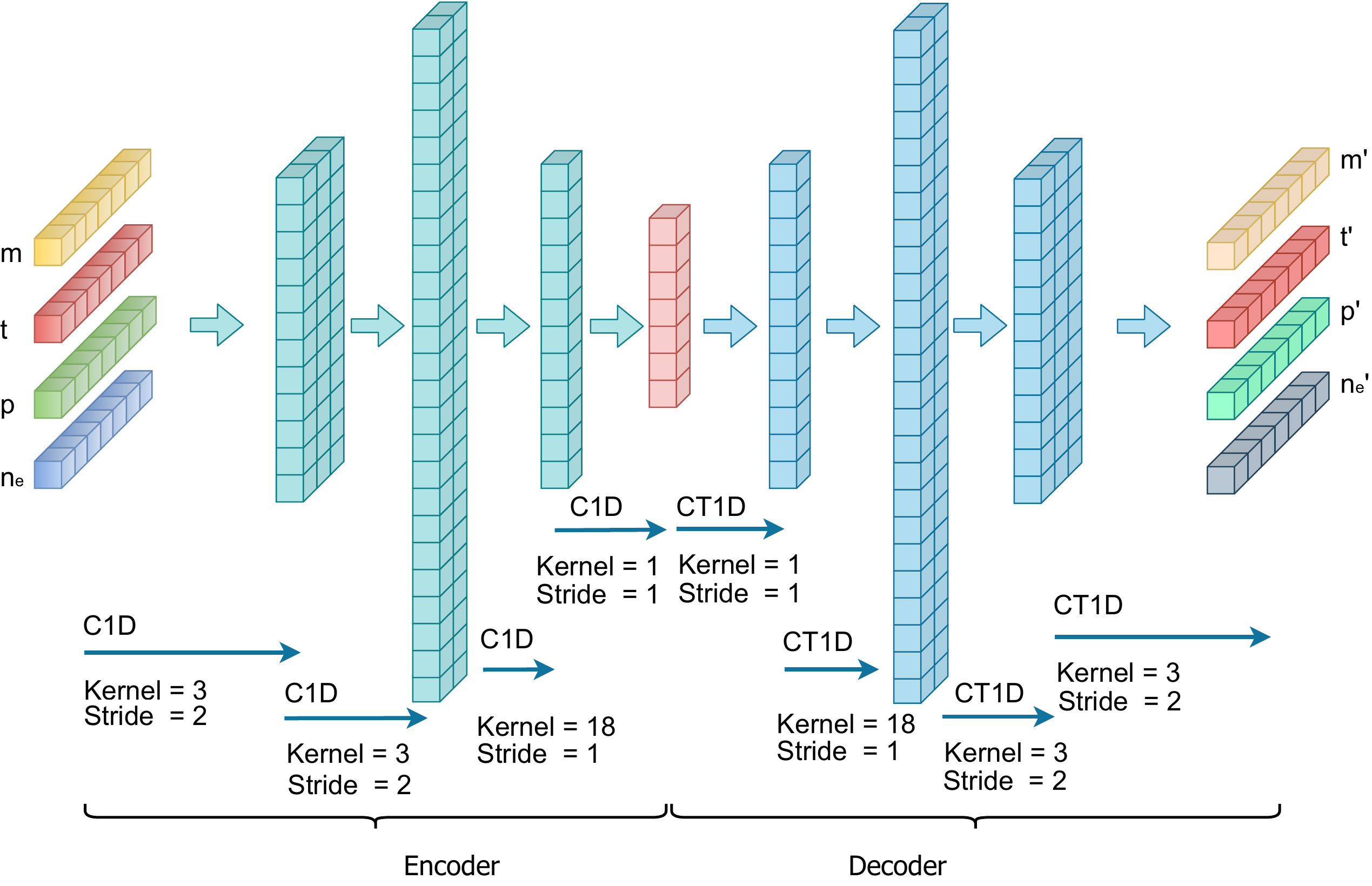}
        \caption{Architecture of our CAE for feature extraction. The 1x1 convolutional layer is at the center to act as the bottleneck. The values obtained at this bottleneck are the result of the encoder part, and constitute the outputs of the NN, the last layer from Fig.~\ref{fig:NN_gslide}. We note that C1D are 1D convolutions and C1TD are 1D transposed convolutions. The activation after each convolutional layer function is always an exponential linear unit (ELU). For convolutional layers the heights were scaled to the number of channels and the depths were scaled to the resulting number of kernels per channel. The detailed code available at  \texttt{https://github.com/cwestend/iNNterpol}}
    \label{fig:CAE_71_b}
\end{figure}

In order to both reduce the dimensions and to extract the nonlinearities in the parameter space, we set up a CAE as depicted in Fig.~\ref{fig:CAE_71_b}. As an AE is essentially a special type of NN in which, in an unsupervised manner, the output mimics the input in the most accurate way, we thus have an equal number of neurons on both input and output layers. The type of CAE we are interested in is a so-called "under-complete" one, where there is a bottleneck layer that has fewer neurons than the input and output ones. To be able to compare it to PCA48-NN results (see Table~\ref{tab:NN}), we first set this bottleneck to the same number of components chosen before (48; 12 are for each of the four physical parameters), then proceeded to increase this number to try to improve our results and to obtain the optimal solution. The bottleneck made the CAE learn the essential features from the input in order to reproduce them accurately. In this way we captured the existing nonlinearities in the model data and go beyond PCA. Also importantly, the bottleneck served as an effective regularizer so the network does not just learn the input values (overfitting) which would hinder its use for data it has not yet seen (i.e., been trained on). We then separated the layers up to the bottleneck and used it as the so-called encoder, the results of which we then fed to the same fully connected NN we used before with PCA. In this way the encoder acts as the feature extractor and dimensionality reducer. Once the NN is trained to predict output values of equal length as the bottleneck given any combination of input parameters, we  then used the decoder part to reconstruct the physical parameters that constitute the predicted stellar atmosphere.  

It is worth noting that in this CAE-NN, the four physical parameters for which we want to extract their features (mass column, temperature, gas pressure, and electronic number density stratification) are fed into the CAE as four channels simultaneously. This means that any relation between these parameters for a specific combination of the grid values (effective temperature, log g, and metalicities) are taken into account combined all together by the NN.

We also found it necessary to normalize the input values of each quantity by subtracting the minimum value of all the models for this quantity and dividing by the difference between the maximum and the minimum of all models. This is common practice in training NNs as the learned weights do not have to span great size differences. We also found that the trained CAE used in the NN that worked best in our case was the exponential linear unit (ELU). The NN that incorporates the bottleneck provided by PCA, on the contrary, performed better using the leaky rectified linear unit. The optimal configurations are summarized in Table~\ref{tab:NN}.

\begin{figure}
        \includegraphics[width=\columnwidth]{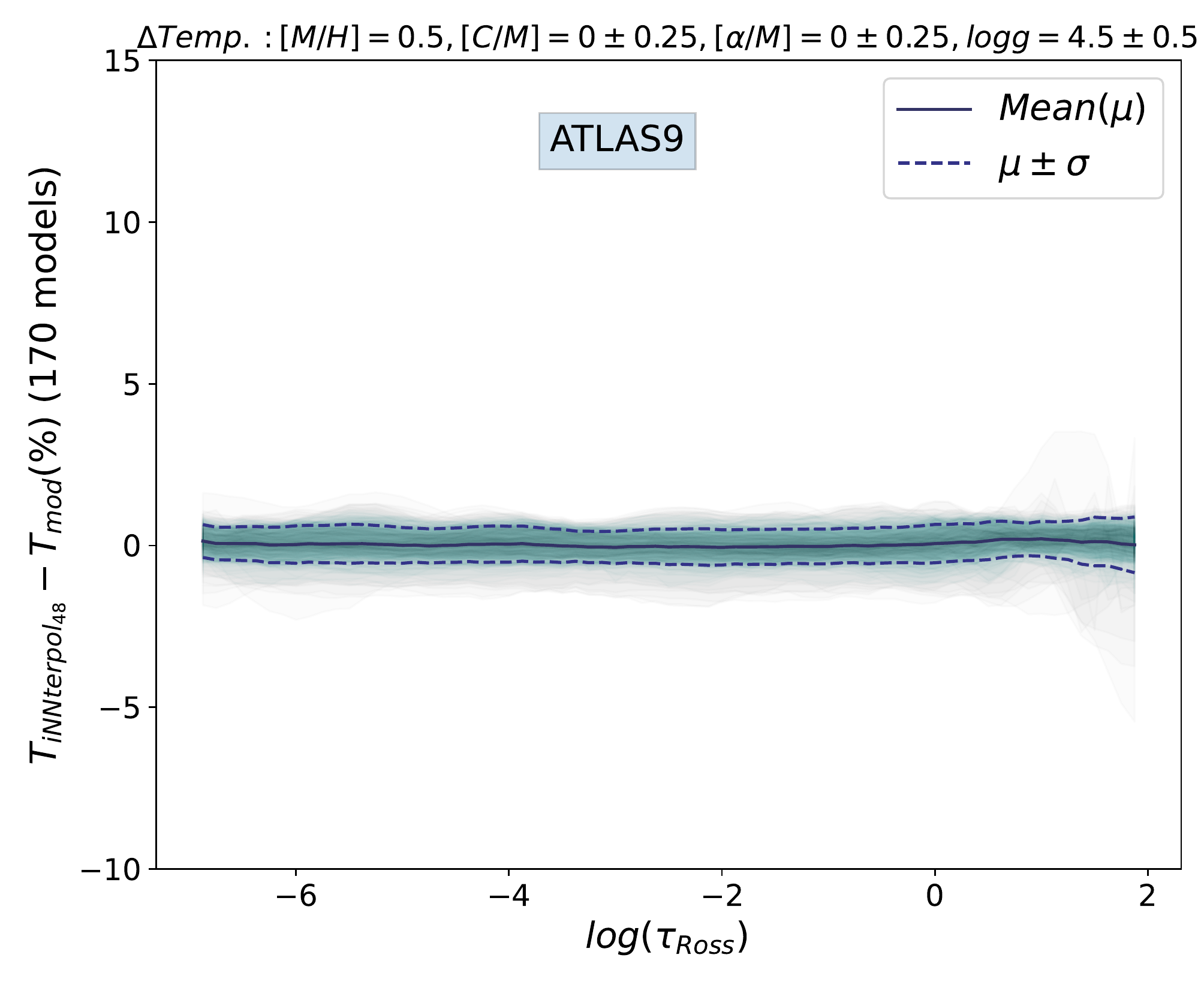}
    \caption{Results for CAE48-NN for ATLAS9 data with a bottleneck of 48 to compare with PCA48-NN (Fig.~\ref{fig:DTemp_chk48_M0_mp05_lgg45}). Differences in temperature as a percentage between the predicted models and the actual ones for points in the grid never seen for the NN. The number of layers chosen were 16 and the number of neurons per layer were 48, instead of 12 layers and 40 neurons per layer we used for PCA. The plot is represented in axis scales so as to compare it with Fig.~\ref{fig:DTemp_chk48_M0_mp05_lgg45}}
    \label{fig:DTemp_chkCAE_16_48_M0_mp05_lgg45}
\end{figure}

\begin{figure}
        \includegraphics[width=\columnwidth]{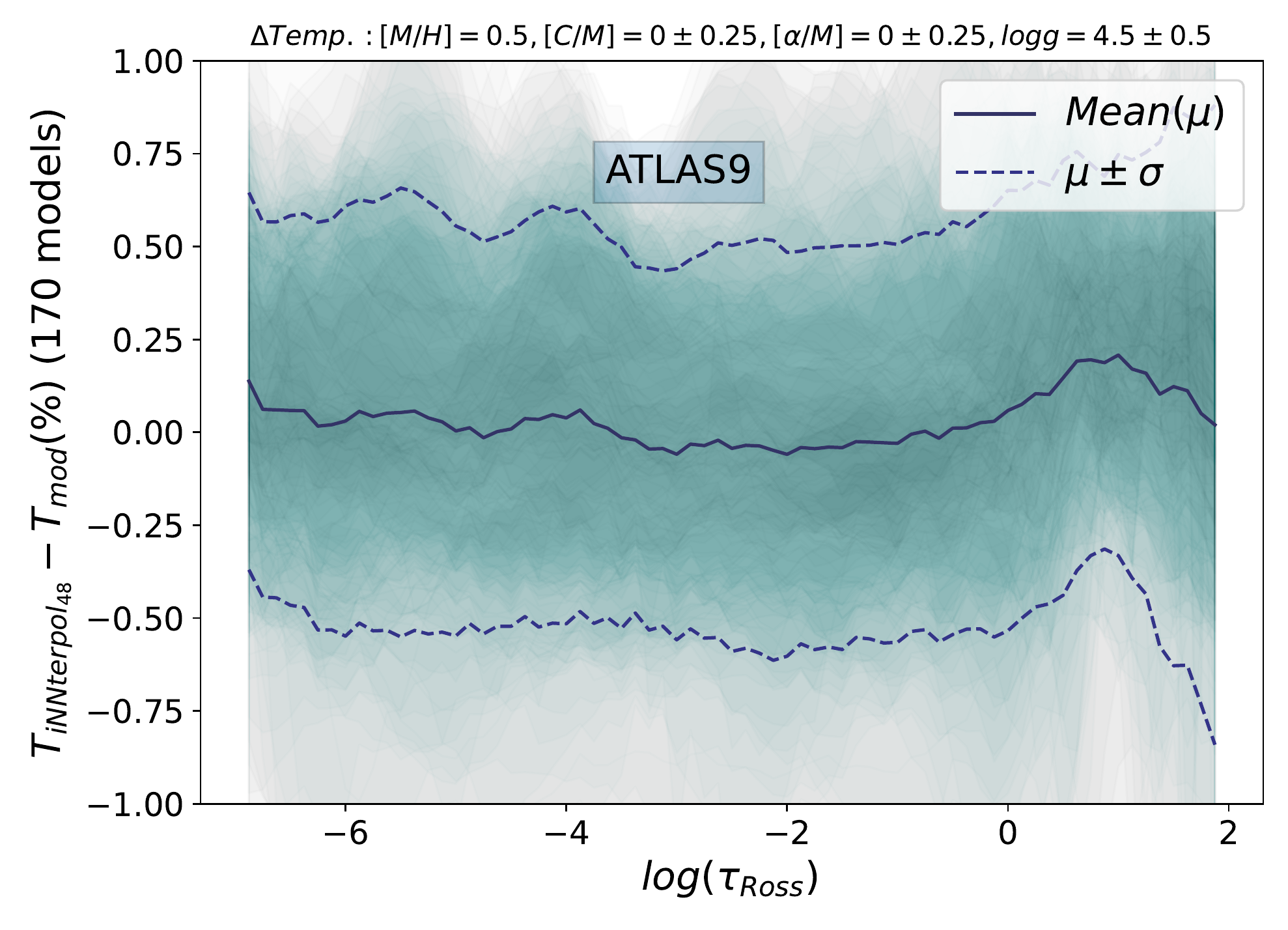}
    \caption{Same as Fig.~\ref{fig:DTemp_chkCAE_16_48_M0_mp05_lgg45}, but rescaled for clarity.}
    \label{fig:DTemp_chkCAE_16_48_M0_mp05_lgg45_scale}
\end{figure}

\begin{figure}
        \includegraphics[width=\columnwidth]{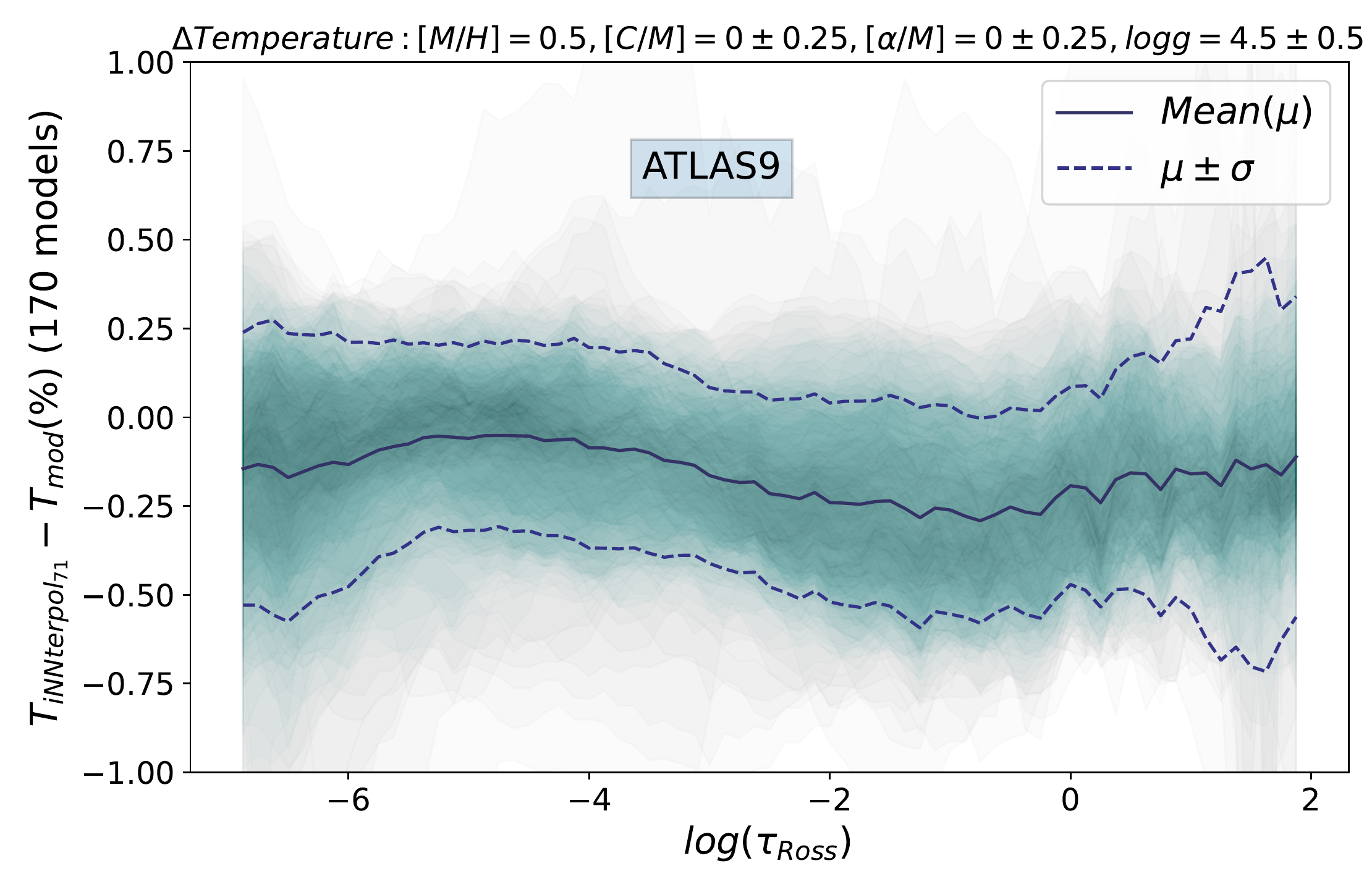}
    \caption{Results of iNNterpol for ATLAS9 data with a bottleneck of 71 and 16 layer deep (CAE71-NN). Shown are differences in temperature in percentage between the predicted models and the actual ground truth ones for points in the grid never seen for the NN. Effective temperatures cover the whole range from 3,500 K to 30,000 K.}
    \label{fig:DTemp_chkCAE_16_71_M0_mp05_lgg45_scale}
\end{figure}

The results of using the CAE with an output of 48 values inside a similar NN (CAE48-NN) as used before in PCA48-NN can be seen in Fig.~\ref{fig:DTemp_chkCAE_16_48_M0_mp05_lgg45}, which is at the same scale as Fig.~\ref{fig:DTemp_chk48_M0_mp05_lgg45} for comparison. In this case we increased the number of layers of the NN to 16 as opposed to 12 for PCA36-NN or PCA48-NN and the number of nodes to 48 instead of 40 (see Table~\ref{tab:NN}). This small increase in free parameters enables the network to converge to a better solution when using the CAE; while using this exact same configuration for PCA, we find a similar solution for the temperatures but then the pressures are poorly fit, underfitting on average by more than 3$\%$. 

The same results are shown in more detail in Fig.~\ref{fig:DTemp_chkCAE_16_48_M0_mp05_lgg45_scale}. The results from CAE48-NN already show an improvement of a factor of about 2 compared to PCA48-NN. We then went even deeper (literally) and gave more power to the NN model, probing the hyperparameter space to obtain a better fit. This was done incrementing the number of total layers from 12 to 16 and the number of nodes per layer from 40 to 71, as well as using a CAE with a bottleneck of 71 values (CAE71-NN). This configuration is what we call the optimal model and shall refer to as iNNterpol. The results are shown in Fig.~\ref{fig:DTemp_chkCAE_16_71_M0_mp05_lgg45_scale} where another doubling of the gain in precision was obtained from the CAE using 48 components and the previous configuration (CAE48-NN, 16 layers, 48 neurons per layer).

Temperature is the best retrieved parameter in all these NNs, both for PCA-NNs and CAE-NNs. In Appendix~\ref{sec:extra} we show the results for mass column, gas pressure, and electronic number density in Fig.~\ref{fig:DMass_chkCAE_16_71_M0_mp05_lgg45_scale}, Fig.~\ref{fig:DPres_chkCAE_16_71_M0_mp05_lgg45_scale}, and Fig.~\ref{fig:DDens_chkCAE_16_71_M0_mp05_lgg45_scale}, respectively.

\begin{figure}
        \includegraphics[width=\columnwidth]{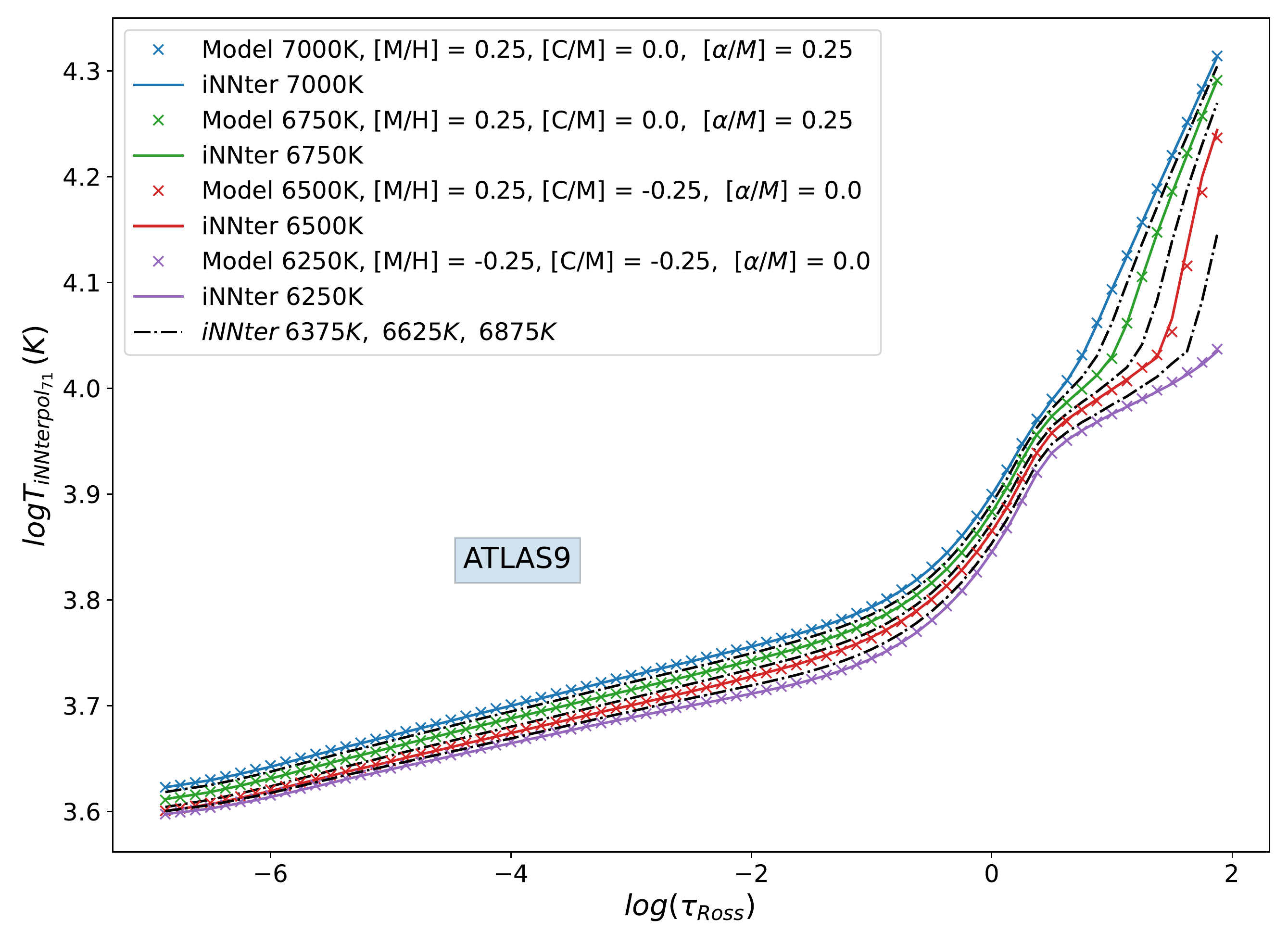}
    \caption{Fits to various values of temperature with iNNterpol with a bottleneck of 71 (CAE71-NN) for ATLAS9 data. For all values, log g = 2.0. The test values not seen by the NN chosen randomly were used, that is why the metallicities are not exactly the same, but are all in an interval of $\pm$0.25 dex. Predicted values are in dash-dotted lines, and we note that interpolation is in temperature and also in metallicities.}
    \label{fig:Temp_CAE_16_71_Teff7000_6250_lgg20}
\end{figure}

The quality of the predictions can be seen in the temperature stratifications shown in Fig.~\ref{fig:Temp_CAE_16_71_Teff7000_6250_lgg20}. These are test models never seen by the NN in training. The metallicities are not identical, as the test models were chosen randomly. They represent a zone of low effective temperature, near-solar abundances, and low surface gravity (all with an identical log g = 2.0), where the calculated models are more critical and where the predictions could most likely fail. We see that the predictions are nevertheless really hard to distinguish from the actual models (sold lines overlap) as the errors in these temperature values are indeed in the 10-20 K range. The exception being at Teff = 6500 K for deep layers ($\tau_{Ross}$ > 1.4) where even the calculated models show a "kink" that is not representative of a realistic stellar atmosphere and it is likely a numerical artifact. The rest of the fits to the other physical quantities are in Appendix~\ref{sec:extra}, Figs.~\ref{fig:Mass_CAE_16_71_Teff7000_6250_lgg20}, \ref{fig:Pres_CAE_16_71_Teff7000_6250_lgg20}, and  \ref{fig:Dens_CAE_16_71_Teff7000_6250_lgg20}.
All cases show that the interpolated values follow the behavior of the grid models in a reasonable way, and we believe it is proof that the iNNterpol NN is able to recover the relations between the parameters and that is reflects it in the obtained models. 

\subsection{Results for MARCS and PHOENIX models}
\label{sec:3models} 

We applied the above CAE configuration to both MARCS and PHOENIX data, parting with the above CAE71-NN (iNNterpol) found for ATLAS9 data. The optimal parameters found are described in Table~\ref{tab:MarcsPhoenix}. The data dimensionality is very different, starting with the number of model atmospheres (853,000 models for ATLAS9, 381,000 for MARCS, and 47,000 for PHOENIX), the number of points in optical depth in the stellar atmosphere (71 for ATLAS9, 56 for MARCS, and 64 for PHOENIX), up to the specific coverage of parameter space. For this reason we tested for an increased range of values, specifically the number of layers $n_l$. To be able to perform this, conscious that increasing the number of fully connected layers can lead to an effective loss of the information in the weights known as the "vanishing gradient problem" (\citealt{Vanishing_2020}, \citealt{Vanishing_2001}) which has been addressed introducing residual networks (ResNet: \citealt{ResNet_2015}), we prevented the loss of information in such a deep NN by making skip connections that allow this information to be carried on. These connections were implemented by bypassing every two fully connected layers. This architecture was useful only for MARCS data, yielding no improvement for ATLAS9 or PHOENIX data. The resulting model is fully detailed online\footnote{\texttt{https://github.com/cwestend/iNNterpol}}.    

\begin{figure}
        \includegraphics[width=\columnwidth]{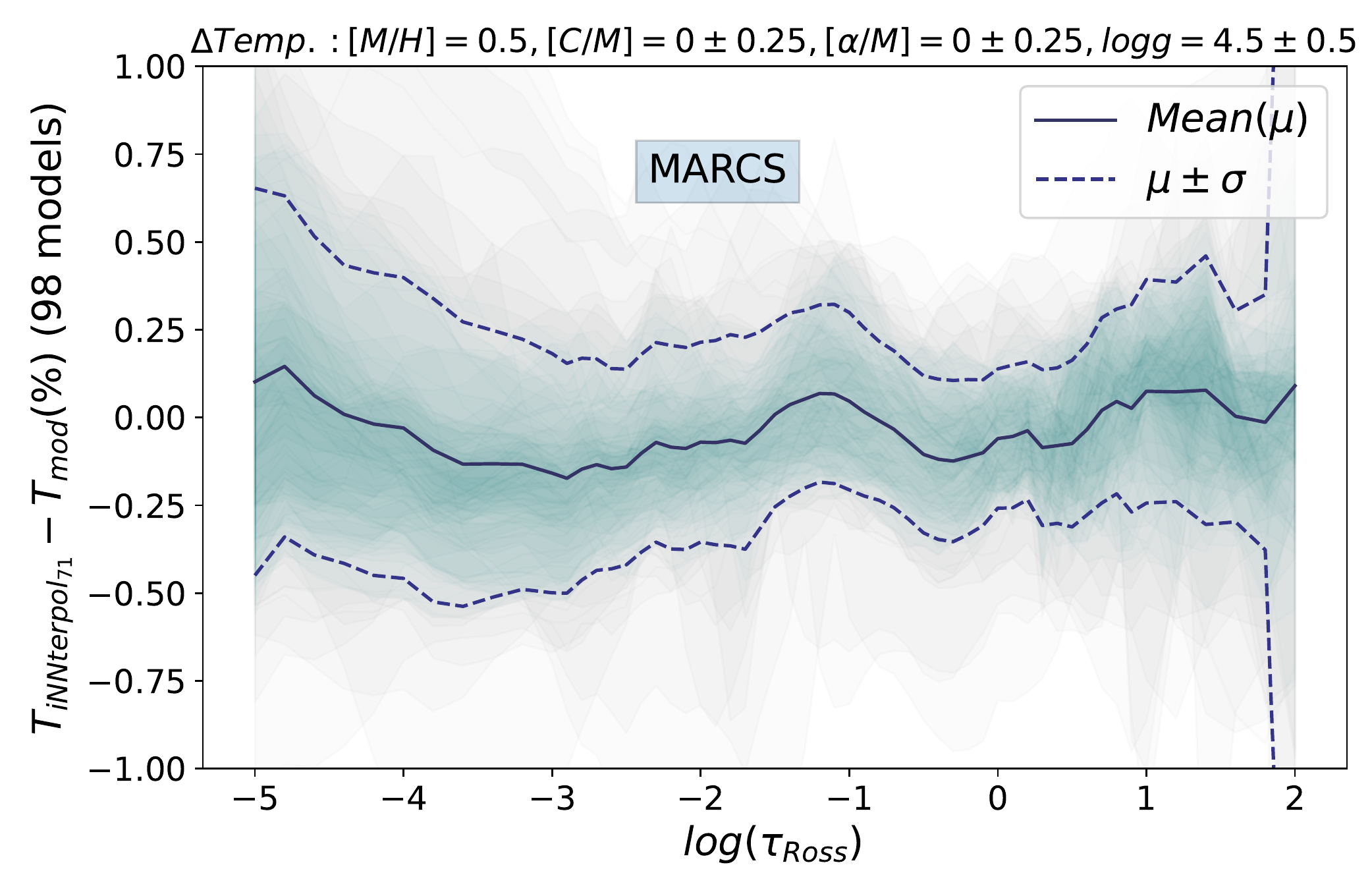}
    \caption{Results of iNNterpol for MARCS data with a bottleneck of 71, 22 layers, and 71 nodes for each layer (CAE71-NN).
    Shown are differences of temperature as a percentage between the predicted models and the actual ground truth ones for points in the grid never seen for the NN. The full range of effective temperatures are covered (2,500 K to 8,000 K).}
    \label{fig:Temp_chkCAE_22_71_resnet_MARCS}
\end{figure}

\begin{figure}
        \includegraphics[width=\columnwidth]{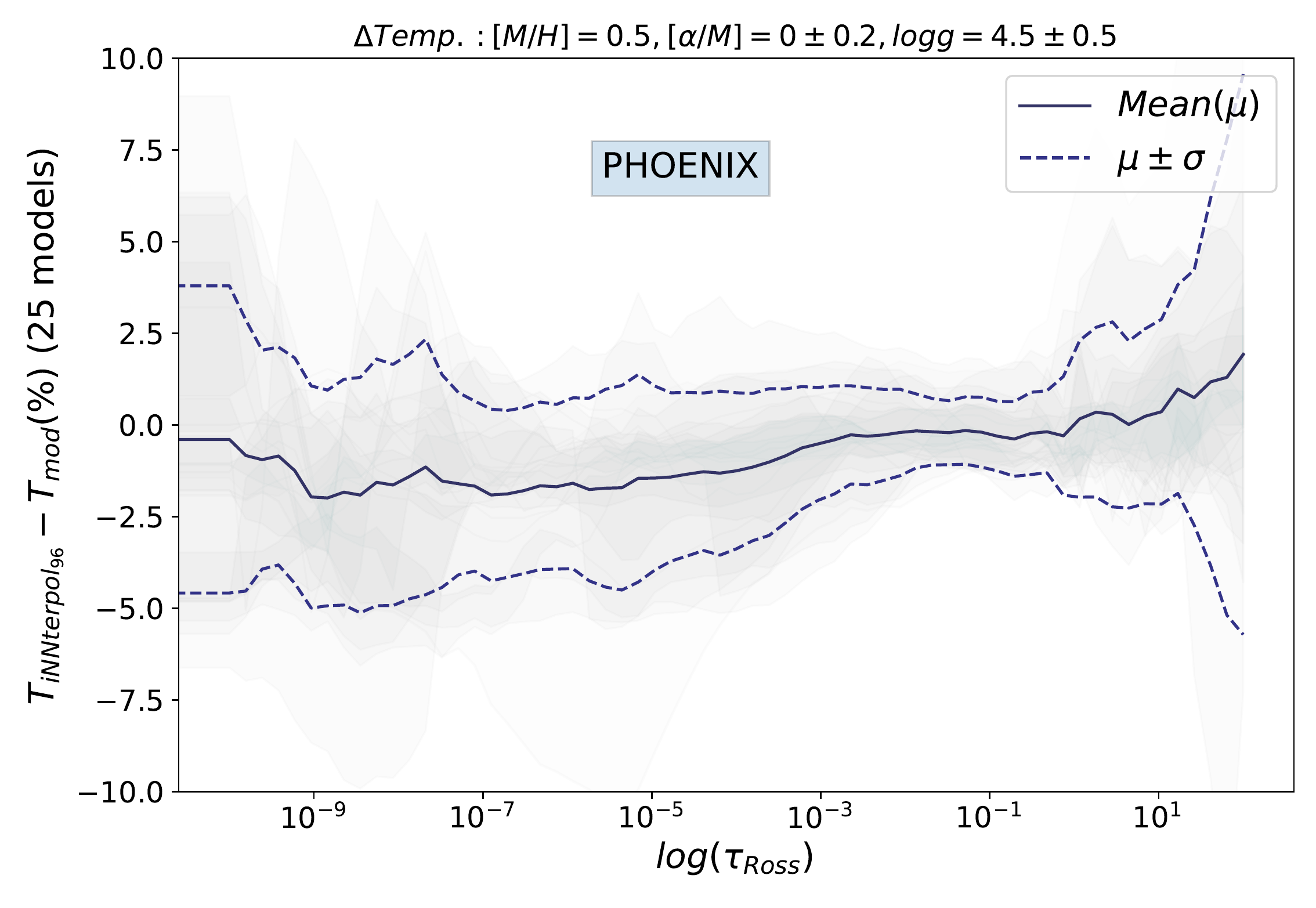}
    \caption{Results of CAE96-NN for PHOENIX data with a bottleneck of 96, 16 layers, and 96 nodes for each layer. Differences in temperature as a percentage between the predicted models and the actual ones for points in the grid never seen for the NN. The full range of effective temperatures are covered (2,500 K to 15,000 K).}
    \label{fig:DTemp_chkCAE_16_96_PHOENIX_10pct}
\end{figure}

The results are illustrated in Fig.~\ref{fig:Temp_chkCAE_22_71_resnet_MARCS} for MARCS data which have an equivalent quality to those for ATLAS9 (Fig.~\ref{fig:DTemp_chkCAE_16_71_M0_mp05_lgg45_scale}), while for PHOENIX the precision is five times worse, as shown in Fig.~\ref{fig:DTemp_chkCAE_16_96_PHOENIX_10pct}.\ We believe this can be explained by the grid with PHOENIX having
an order of magnitude less models, being much less dense as compared to those of ATLAS9 and MARCS.

\begin{table}
    \caption{Range of parameters tested for the fully connected NN and optimal values for MARCS and PHOENIX data. For both MARCS and PHOENIX, the best configurations have the resulting dimensionality reduction (n$_o$) equal to the neurons per layer (n$_n$) which are n$_o$ = n$_l$ = 71 and 96, respectively.}
    \centering
    \label{tab:MarcsPhoenix}
    \small
        \begin{tabular}{cccc} 
                \hline
                n$_l$ Layers & n$_n$ Neurons/layer & Activation & Batch size\\
                \hline
                16 to 28 & 64 to 128 & lkrelu, elu & 128, 64, 32\\
                \hline
            Optimal values\\
            \hline
                (MARCS CAE71)\\
                22 & 71 & elu & 64\\
                (PHOENIX CAE96)\\
                16 & 96 & elu & 64 \\
                \hline
        \end{tabular}
\end{table}

\subsection{Comparison to linear interpolation}
\label{sec:linear} 

In the specialized literature, it is common practice to interpolate model atmospheres linearly for the desired atmospheric parameters. Although models could be computed afresh for those values, the codes are not always public or easy to use, while precomputed model grids are available.
Therefore, an interesting question is what the differences are between using classical linear interpolation and our iNNterpol method on the parameter space. If the grid is sufficiently dense in the sense that the models are close enough together, and thus vary linearly from one to the next, linear interpolation should be a good resort, but this is very hard to know in advance. Furthermore, there are places where a slight variation of even a single parameter can yield a very different model as we discuss below. 

The problem is further enhanced in places where this grid is not dense enough as near the grid limits or edges, or where there are missing models due to a lack of convergence. To test the validity of our method, we proceeded to create a new grid by linearly interpolating at half-step intervals on the original one. Using this new grid, we could use it again at half-step intervals to be able to compare the calculated values to the original ones used to make the new grid in the first place as described in \citet{Sara_grid_test}. Making the assumption that the errors in this way are independent and Gaussian, the resulting deviations have to be corrected by a factor of $\sqrt{2}$. We applied this to the exact same atmospheric models used to test our iNNterpol method, specifically those in which all surrounding models exist and a linear 5D interpolation was in fact possible and did not lead to extrapolation. Due to the fact that to be able to perform a linear 5D interpolation, for each desired interpolated atmosphere all 32 ($2^5$) models around the desired value must exist, the number of models available was greatly reduced.

\begin{figure}
        \includegraphics[width=\columnwidth]{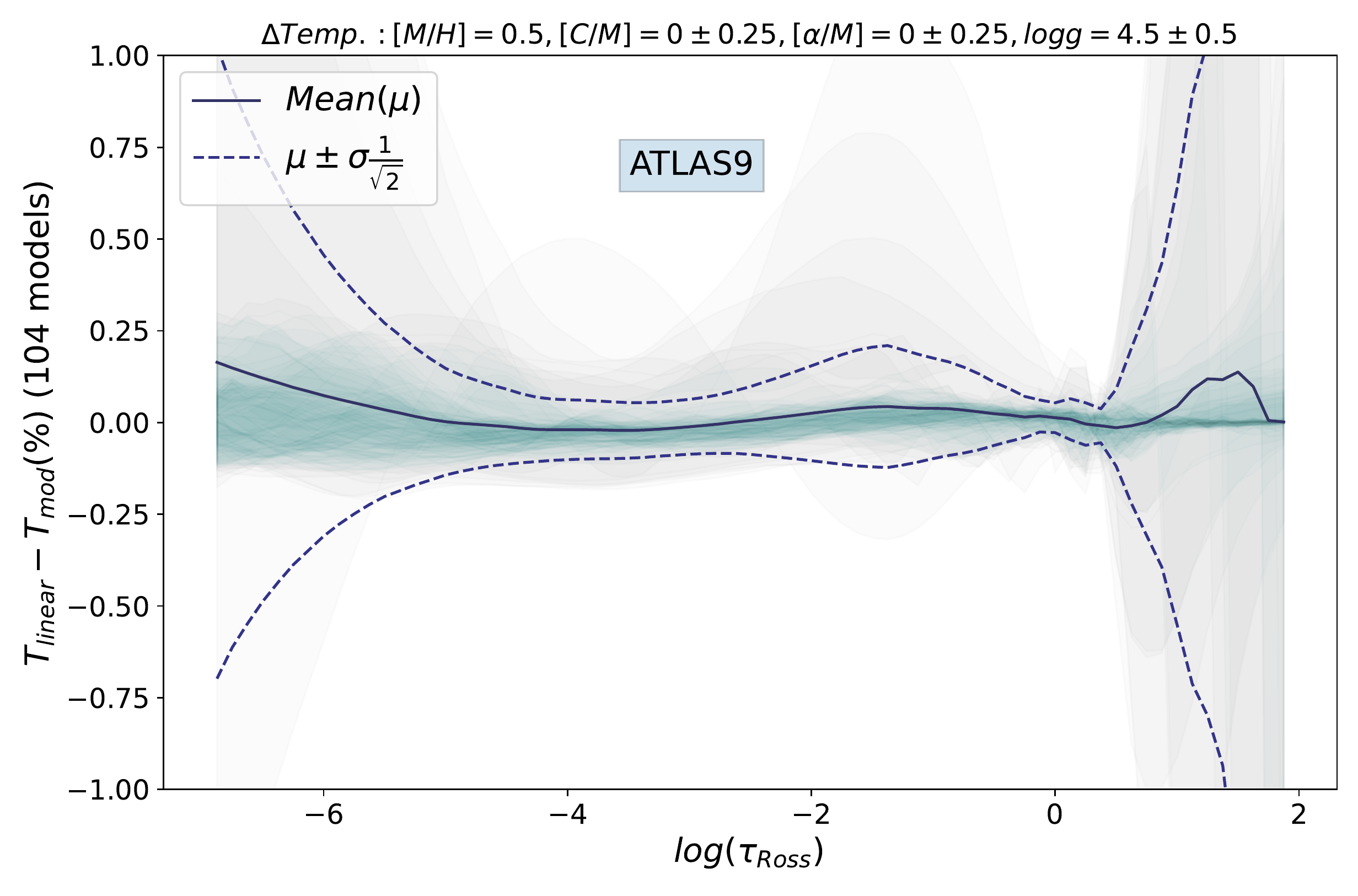}
    \caption{Resulting error of linearly interpolating in a new half-step subgrid of ATLAS9 to compare with the original model atmospheres used for making this new subgrid. Chosen points are the exact same test values used to evaluate iNNterpol (a subset of those shown in  Fig.~\ref{fig:DTemp_chkCAE_16_71_M0_mp05_lgg45_scale}) where actual linear interpolation is possible. The same range applies of Teff from 3,500 K to 30,000 K.}
    \label{fig:DTemp_Linear_ATLAS9_mp05_lgg40_lgg50_1pct}
\end{figure}

\begin{figure}
        \includegraphics[width=\columnwidth]{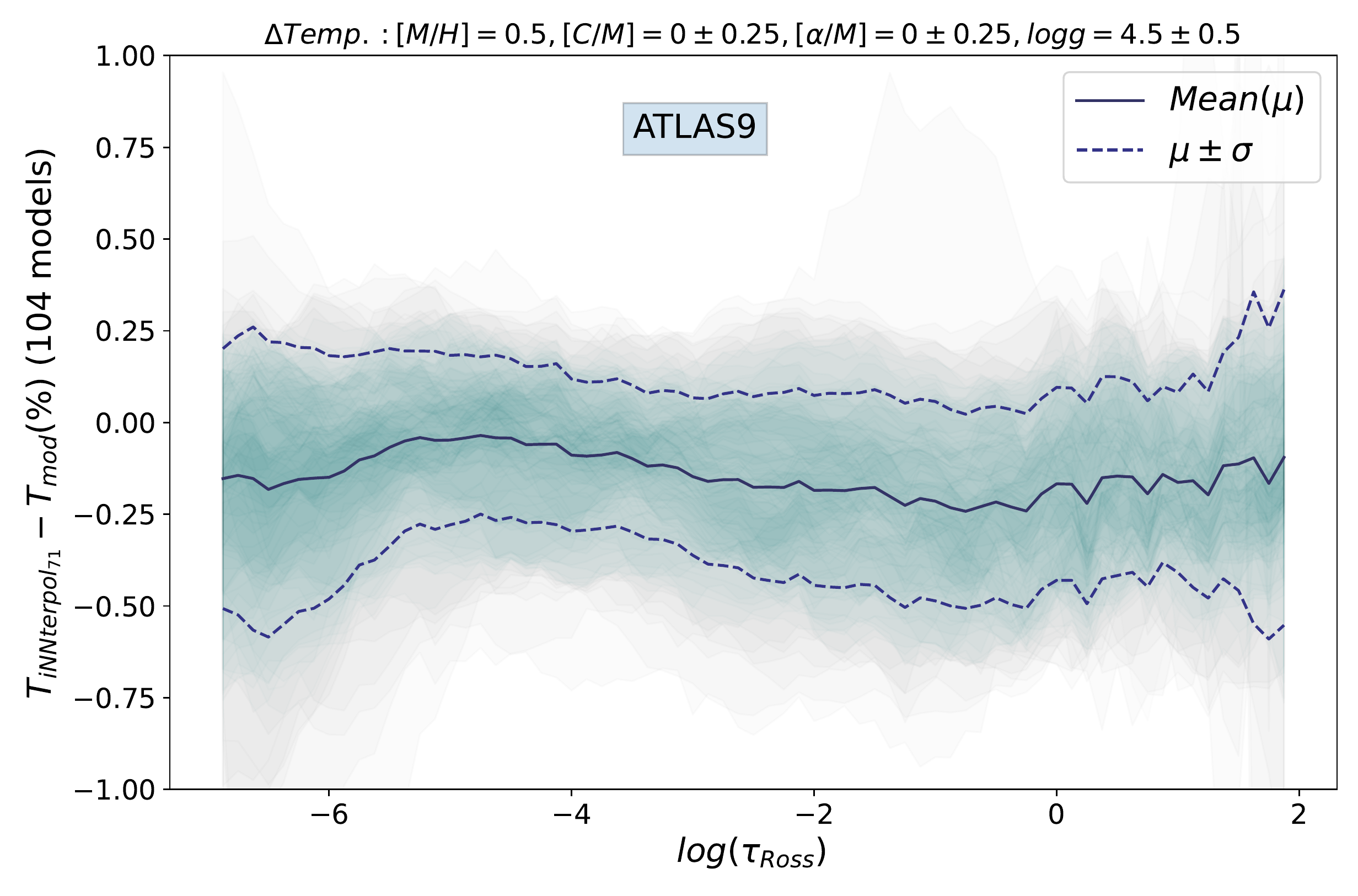}
    \caption{iNNterpol (CAE71-NN) for ATLAS9 for the same models as in Fig.~\ref{fig:DTemp_Linear_ATLAS9_mp05_lgg40_lgg50_1pct} for direct comparison.}.
    \label{fig:DTemp_chkCAE_16_71_ATLAS_mp05_lgg45_1pct}
\end{figure}

\begin{figure}
        \includegraphics[width=\columnwidth]{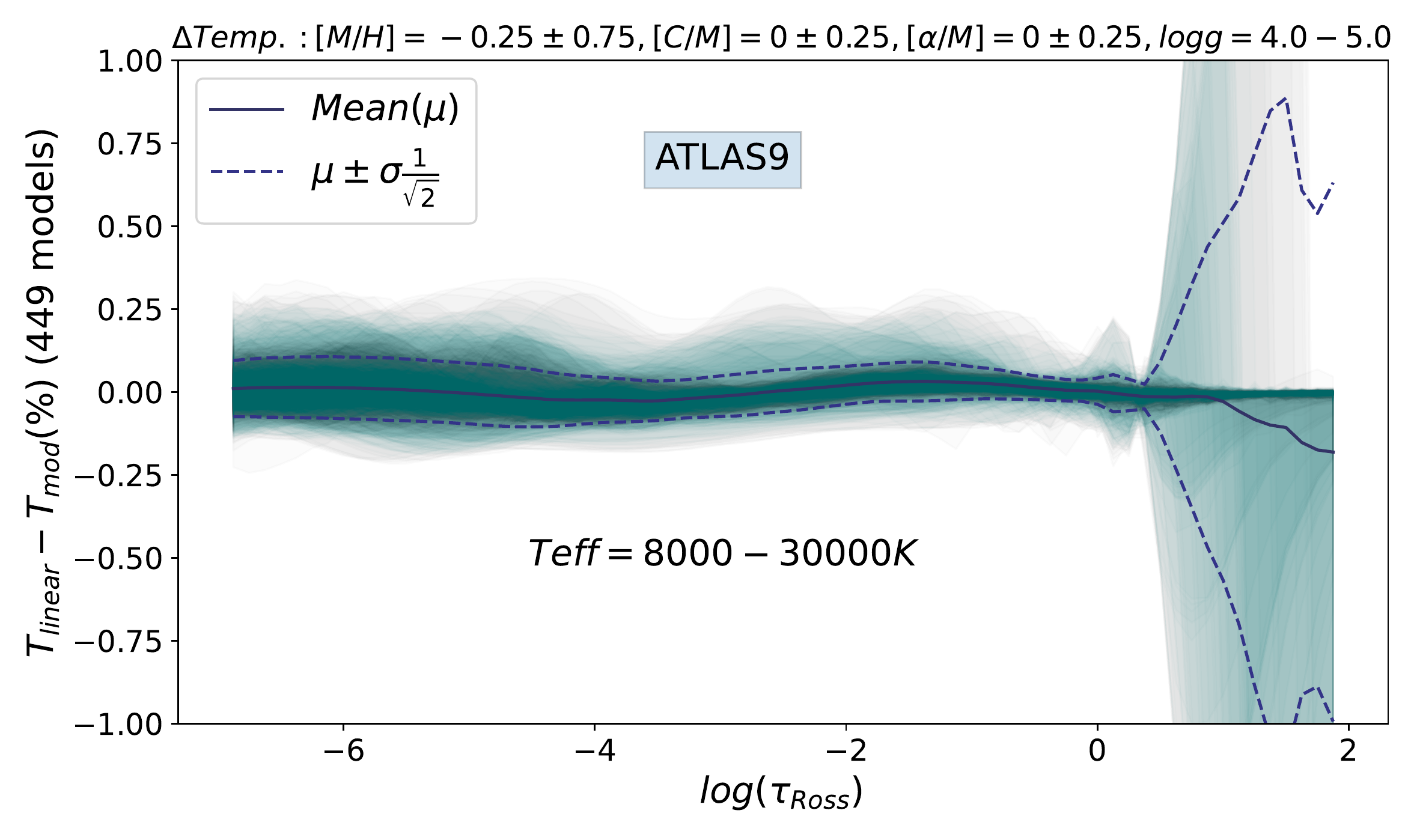}
    \caption{Errors of linearly interpolating on a half-step grid for ATLAS9 hot dwarf stars (Teff between 8,000 K and 30,000 K).}
    \label{fig:DTemp_Linear_ATLAS9_mn10_mp05_lgg40_lgg50_Teff8k_Teff30k_hot_1pct}
\end{figure}

\begin{figure}
        \includegraphics[width=\columnwidth]{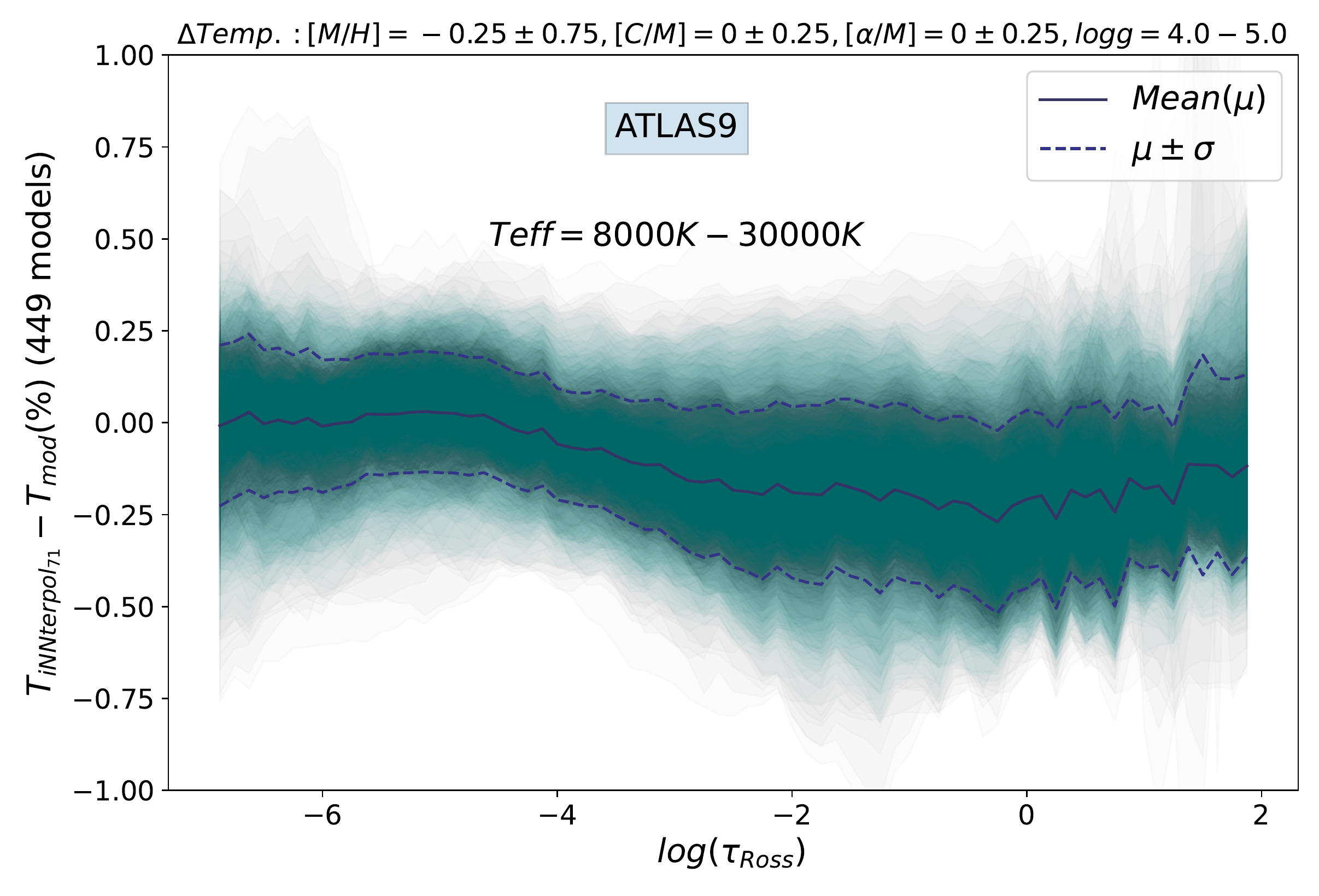}
    \caption{Errors of iNNterpol (CAE71-NN) for the same models as in Fig.~\ref{fig:DTemp_Linear_ATLAS9_mn10_mp05_lgg40_lgg50_Teff8k_Teff30k_hot_1pct} for direct comparison.}
    \label{fig:}
\end{figure}

The errors in linear interpolation are shown in  Fig.~\ref{fig:DTemp_Linear_ATLAS9_mp05_lgg40_lgg50_1pct} where of the initial 170 test models shown in Fig.~\ref{fig:DTemp_chkCAE_16_71_M0_mp05_lgg45_scale}, only 104 remain that have all surrounding model atmospheres to permit a linear interpolation. We note that linear interpolation is very precise for layers at intermediate depths, but the deviations at deep optical depths or high in the stellar atmospheres make this technique less reliable as the RMS deviations become large. Comparing iNNterpol errors over the same grid points as can be seen in  Fig.~\ref{fig:DTemp_chkCAE_16_71_ATLAS_mp05_lgg45_1pct} presents a more constant deviation all through the model atmospheres. 

To evaluate these errors on a different value range, we can for example fix the effective temperature range to that of hot stars for the same (log) gravity range, and this is shown in Fig.~\ref{fig:DTemp_Linear_ATLAS9_mn10_mp05_lgg40_lgg50_Teff8k_Teff30k_hot_1pct}. We see that while at higher layers the error is very small, models at deep layers present a wider variability and thus large error in linear interpolation. Again, our iNNterpol method is more consistent as it gives similar errors throughout the depth scale of the stellar atmosphere. The same behavior is evidenced if we go to other value ranges, such as for cool dwarfs which is shown in Fig.~\ref{fig:DTemp_Linear_ATLAS9_mn10_mp05_lgg40_lgg50_Teff35k_Teff8k_cold_1pct}  and Fig.~\ref{fig:DTemp_chkCAE_16_71_ATLAS_mn10_mp05_logg4_logg5_Teff35k_Teff8k_cold_1pct} or for values for giant stars as shown in Fig.~\ref{fig:DTemp_Linear_ATLAS9_mn10_mp05_lgg10_lgg30_Teff4k_Teff6k_giants_1pct} and Fig.~\ref{fig:DTemp_chkCAE_16_71_ATLAS_mn10_mp05_logg10_logg30_Teff4k_Teff6k_giants_1pct}, contained in the Appendix~\ref{sec:extra}.

\begin{figure}
        \includegraphics[width=\columnwidth]{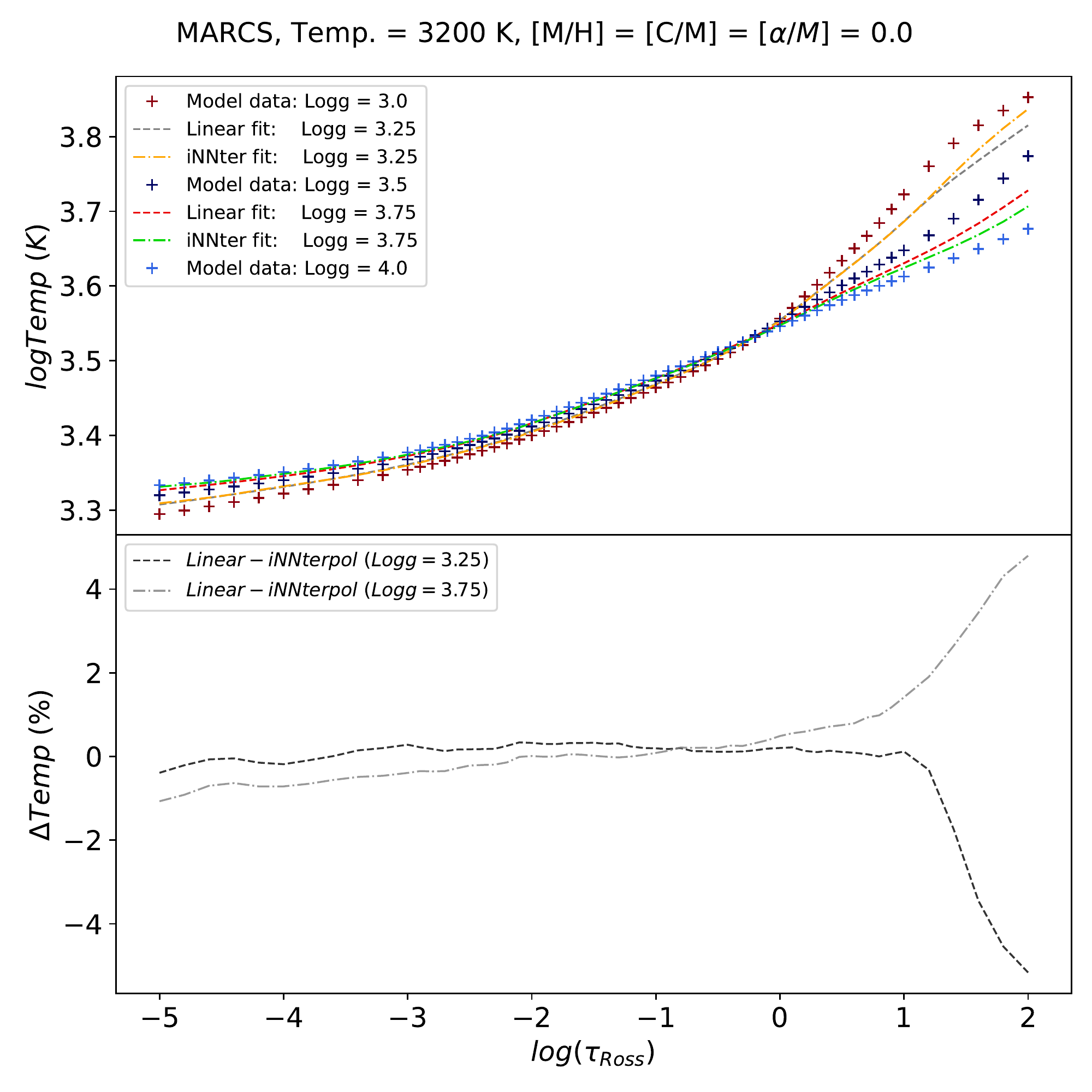}
    \caption{Resulting linear and iNNterpol models fits for MARCS atmospheres with fixed effective temperature and metallicities but varying intermediate log g. Lower panel: Differences between both linear and iNNterpol fits.}
    \label{fig:Temp_Linear_CAE_16_71_Teff3200_Logg3_Logg4_Jump}
\end{figure}

To test how our iNNterpol performs differently to a 5D linear interpolation, we can sample values where the models change for a small step in one parameter. This is shown in Fig.~\ref{fig:Temp_Linear_CAE_16_71_Teff3200_Logg3_Logg4_Jump} for MARCS model atmospheres where all parameters except (log) gravity are kept constant (Teff = 3,200K, [M/H] = [C/M] = [$\alpha$/M] = 0). The interpolated models for intermediate gravity values show that both a linear fit and iNNterpol offer significantly different atmospheres, as the differences between both interpolations are well over the RMS errors previously shown. The effect is more significant at lower depths for both log g values, and also throughout the atmosphere as is the case for log g = 3.75. This region is a specially interesting one as log g = 3.5 marks the transition from spherical models to plane-parallel atmospheres in MARCS models. Here both techniques should naturally differ as linear interpolation takes into account only the contiguous models, while iNNterpol captures information from the whole grid. 

\begin{figure}
        \includegraphics[width=\columnwidth]{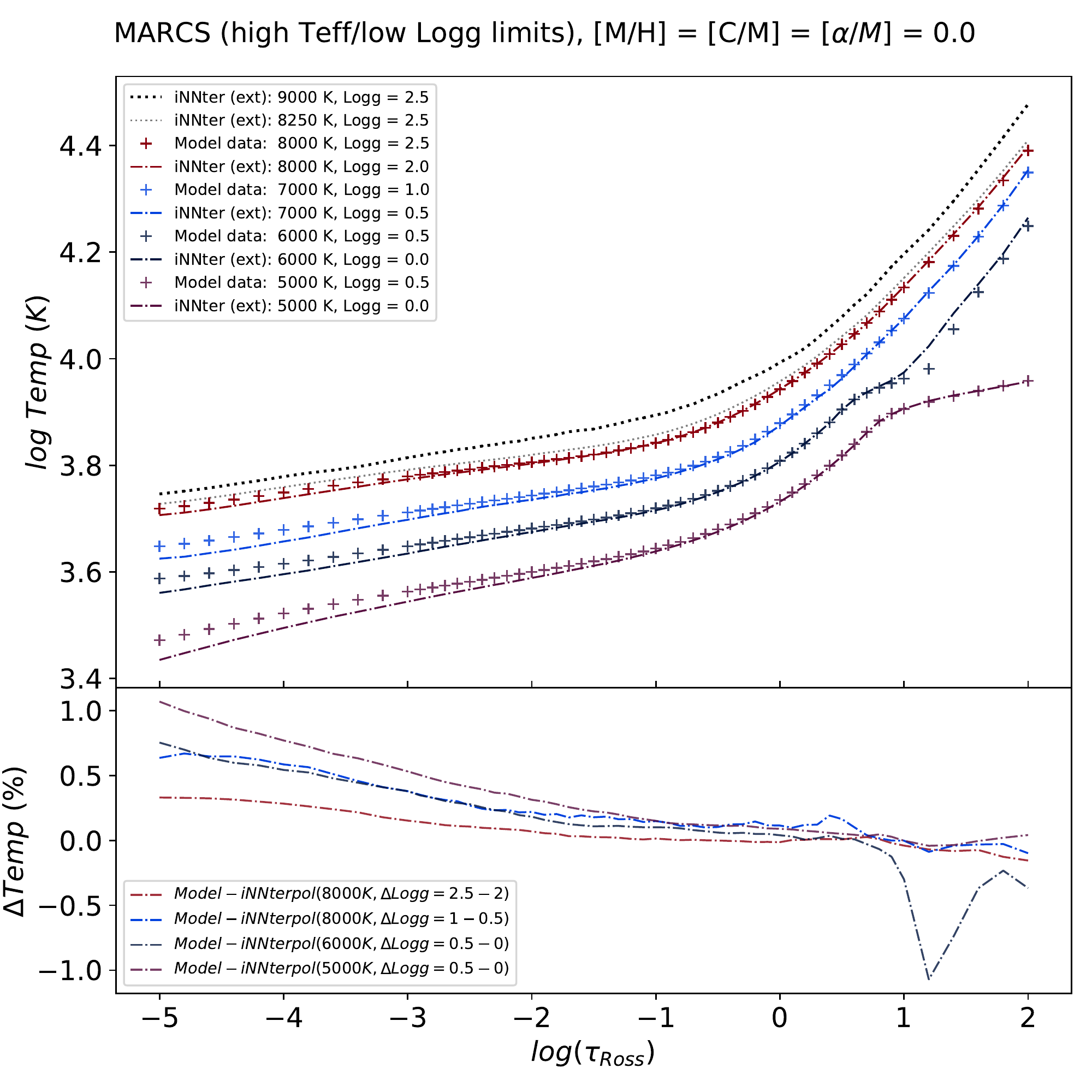}
    \caption{iNNterpol extrapolation models for MARCS atmospheres outside the available grid. Lower panel: Differences between model data and iNNterpol generated extrapolations.}
    \label{fig:Temp_CAE_16_71_Teff8500_5000_Limits}
\end{figure}

One interesting aspect is to be able to go beyond where 5D linear interpolation is not available as there are basically no surrounding models, but our iNNterpol method can still give interesting results. These have to be taken with caution, but providing the jump in a parameter in not too far away; the results may show what the NN is learning about the latent space of model atmospheres as it not only learns from nearby grid points but from all available models. This is shown in Fig.~\ref{fig:Temp_CAE_16_71_Teff8500_5000_Limits} where we extrapolated to find models that are off the grid such as those with high effective temperatures and low gravities. The model atmosphere is the limit in the sense that for MARCS atmospheres, there are no existing models for those Teff beyond the shown (log) gravity values. A model that extrapolates to both Teff = 8,250K and log g = 2,5 is also represented (one grid step beyond the edge of each Teff and log g), and it can be considered as a meaningful limit. Going beyond this (i.e., higher Teff or lower log g or both) yields nonsmooth temperature stratifications as is clearly seen in the model with Teff = 9,000K and log g = 2.5. We consider these variations or inhomogeneities to be artifacts due to the NN not being able to retrieve information for parameter values so far away from the available grid. Our iNNterpol method has the added advantages of being extremely fast and lightweight, as all NN weights and code are under 20MB, while a linear interpolator needs all models which for the MARCS grid take around 1GB of data.

\section{Conclusions}

We hereby present a method for effectively extracting the nonlinearities in a grid of model atmospheres and being able to recover them for any values of the parameter space with great precision. This tool not only is an extremely fast and lightweight way of working with these stellar models, but it demonstrates a technique that can be employed with other families of models, provided the grid covered by these models in parameter space is dense enough. We provide iNNterpol, a fast and reliable interpolator for the ATLAS9 and MARCS family of models. We have shown that using the encoder from a CAE inside our NN greatly increases the power to interpret the existing nonlinearities present in the data, and which surpasses PCA in this configuration. In this way, traditional ML learning techniques, such as LightGBM which are way faster in many cases here, fail to capture the specific variations of the data at hand. We also show our iNNterpol method provides meaningful information where the variation between atmospheres in the grid depart from linearity.

The effort designing and implementing this specific NN with a CAE for feature extraction should serve as a staring point for others in improving and using it for other means and with data of a similar nature, which is abundant in the natural sciences. The full code and the data are freely available in the tradition of deep learning so that others will improve on it without having to recreate its results as all the details are available and open-sourced.

\begin{acknowledgements}
We want to appreciate the useful discussions with Thomas Masseron on linear interpolation and also Ivana Escala for her implementation of Masseron's code (\texttt{https://github.com/iaescala/interp-marcs/}).  
We would also like to thanks our referee, Mikhail Kovalev, whose insightful comments have helped us improve the clarity of this work. CAP acknowledges financial support from the Spanish Ministry of Science and Innovation (MICINN) project PID2020-117493GB-I00. We acknowledge financial support from the Spanish Ministerio de Ciencia, Innovación y Universidades through project PGC2018-102108-B-I00 and FEDER funds.
\end{acknowledgements}

%
%

\bibliographystyle{aa} 
\bibliography{arXiv_corr_final} 



\begin{appendix}
\section{Supplementary material}
\label{sec:extra} 

\begin{figure}
        \includegraphics[width=\columnwidth]{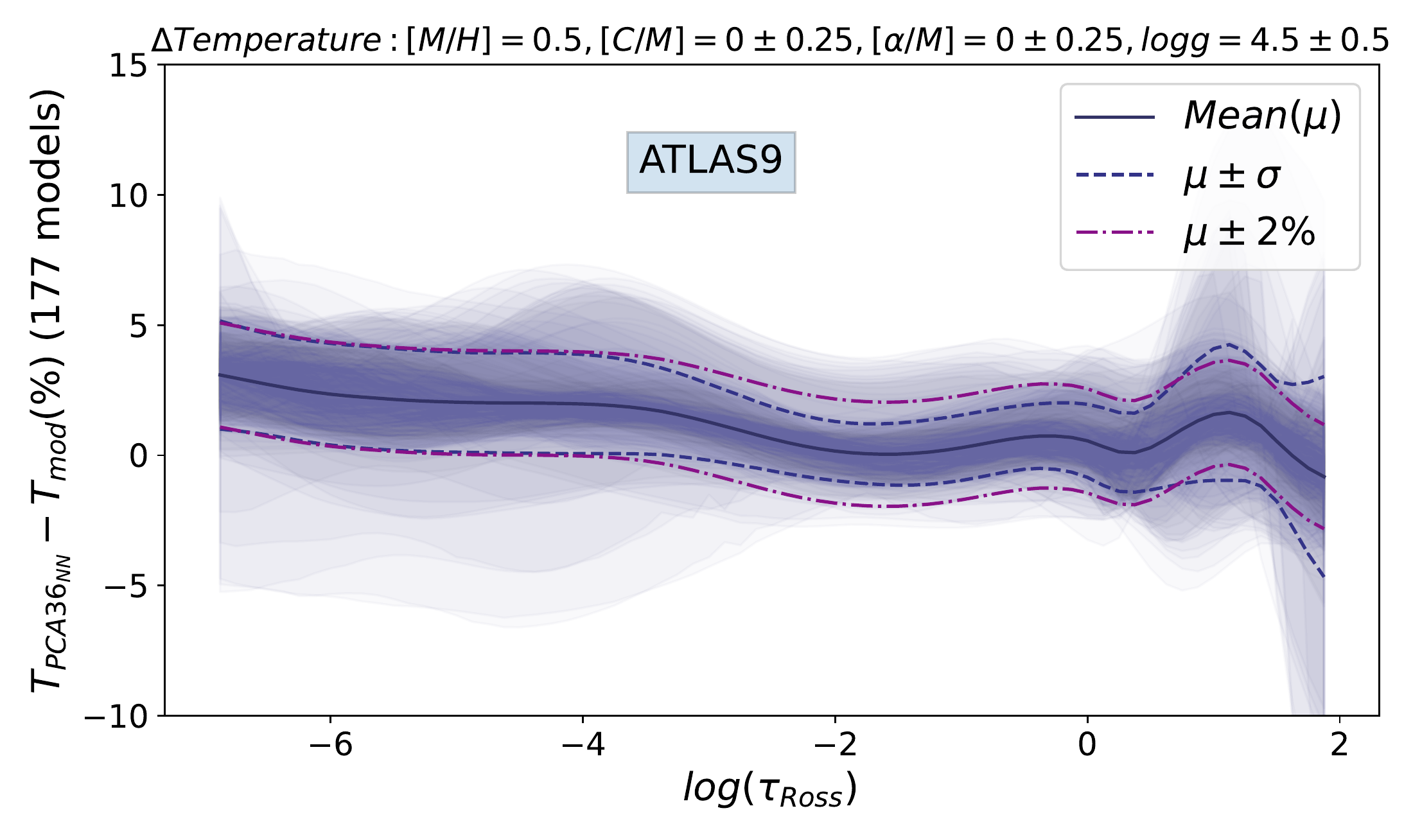}
    \caption{PCA36-NN (nine component PCA) differences in temperature as a percentage between the predicted models and the actual ones for points in the grid never seen for the NN for ATLAS9 model data.}
    \label{fig:DTemp_chk36_M0_mp05_lgg45}
\end{figure}

\begin{figure}
        \includegraphics[width=\columnwidth]{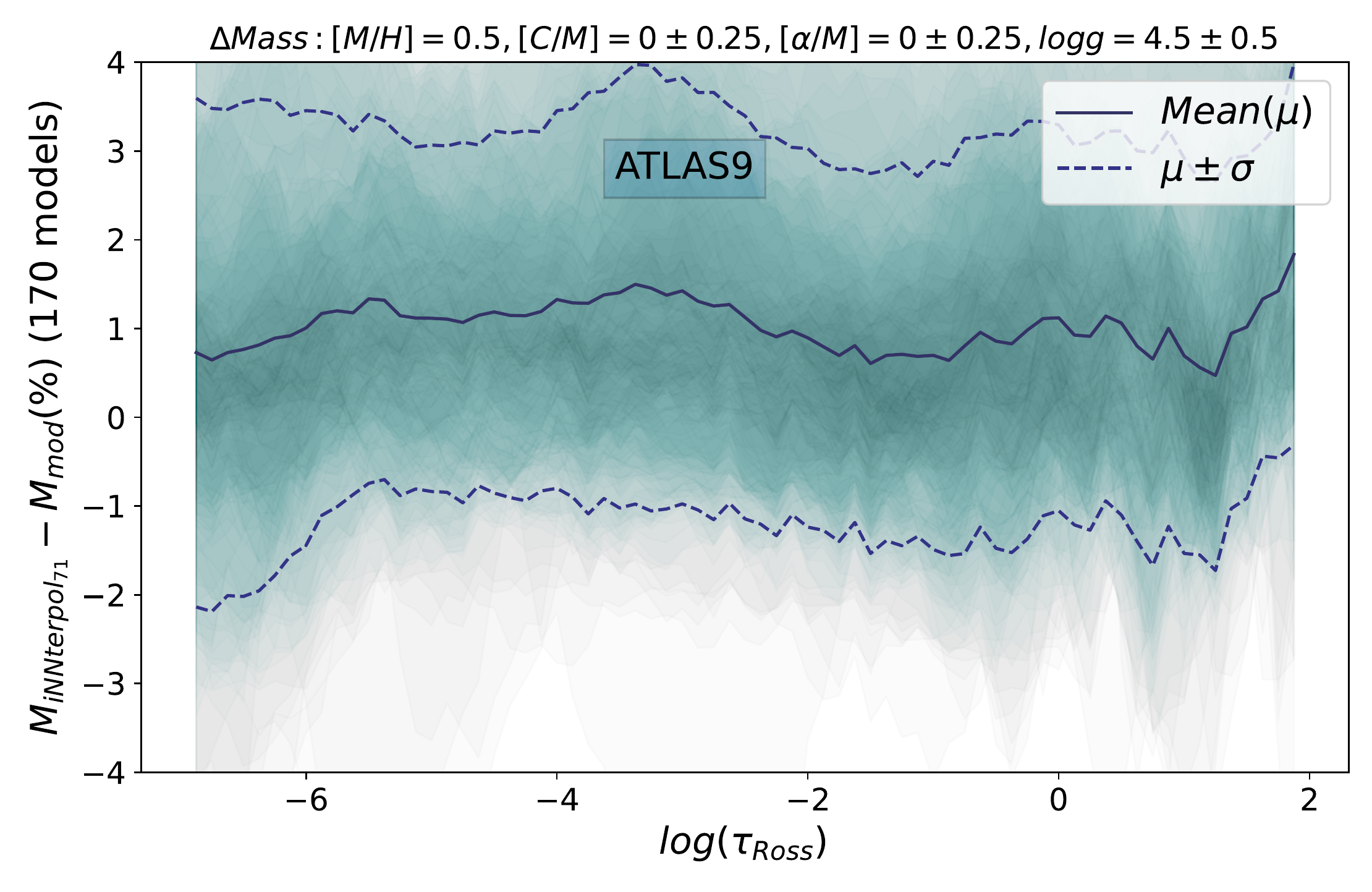}
    \caption{iNNterpol with a bottleneck of 71 (CAE71-NN) differences for mass column density as a percentage between the predicted models and the actual ones for points in the grid never seen for the NN for ATLAS9 model data.}
    \label{fig:DMass_chkCAE_16_71_M0_mp05_lgg45_scale}
\end{figure}

\begin{figure}
        \includegraphics[width=\columnwidth]{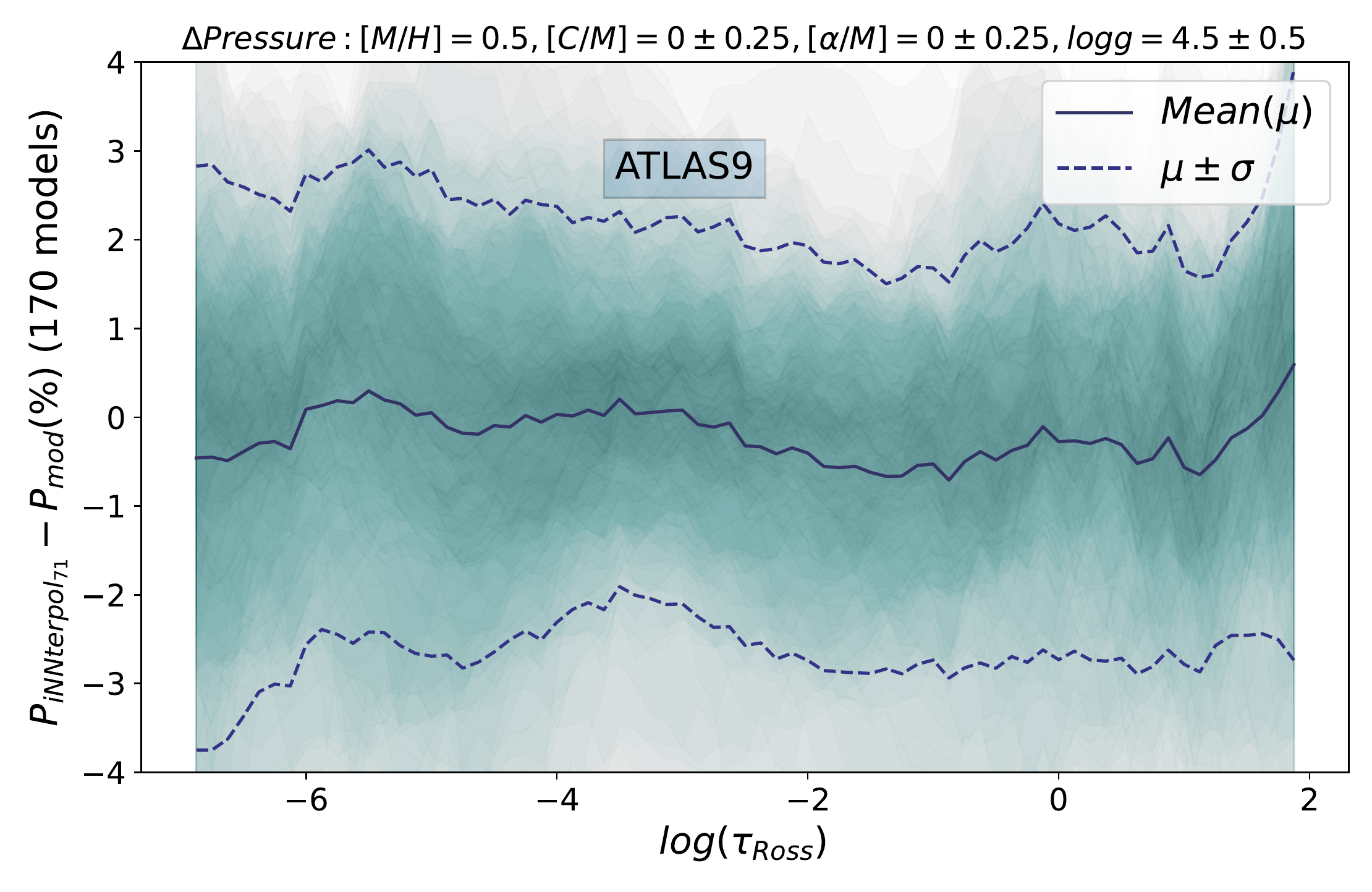}
\caption{iNNterpol with a bottleneck of 71 (CAE71-NN) differences for gas pressure as a percentage between the predicted models and the actual ones for points in the grid never seen for the NN for ATLAS9 model data.}
    \label{fig:DPres_chkCAE_16_71_M0_mp05_lgg45_scale}
\end{figure}

\begin{figure}
        \includegraphics[width=\columnwidth]{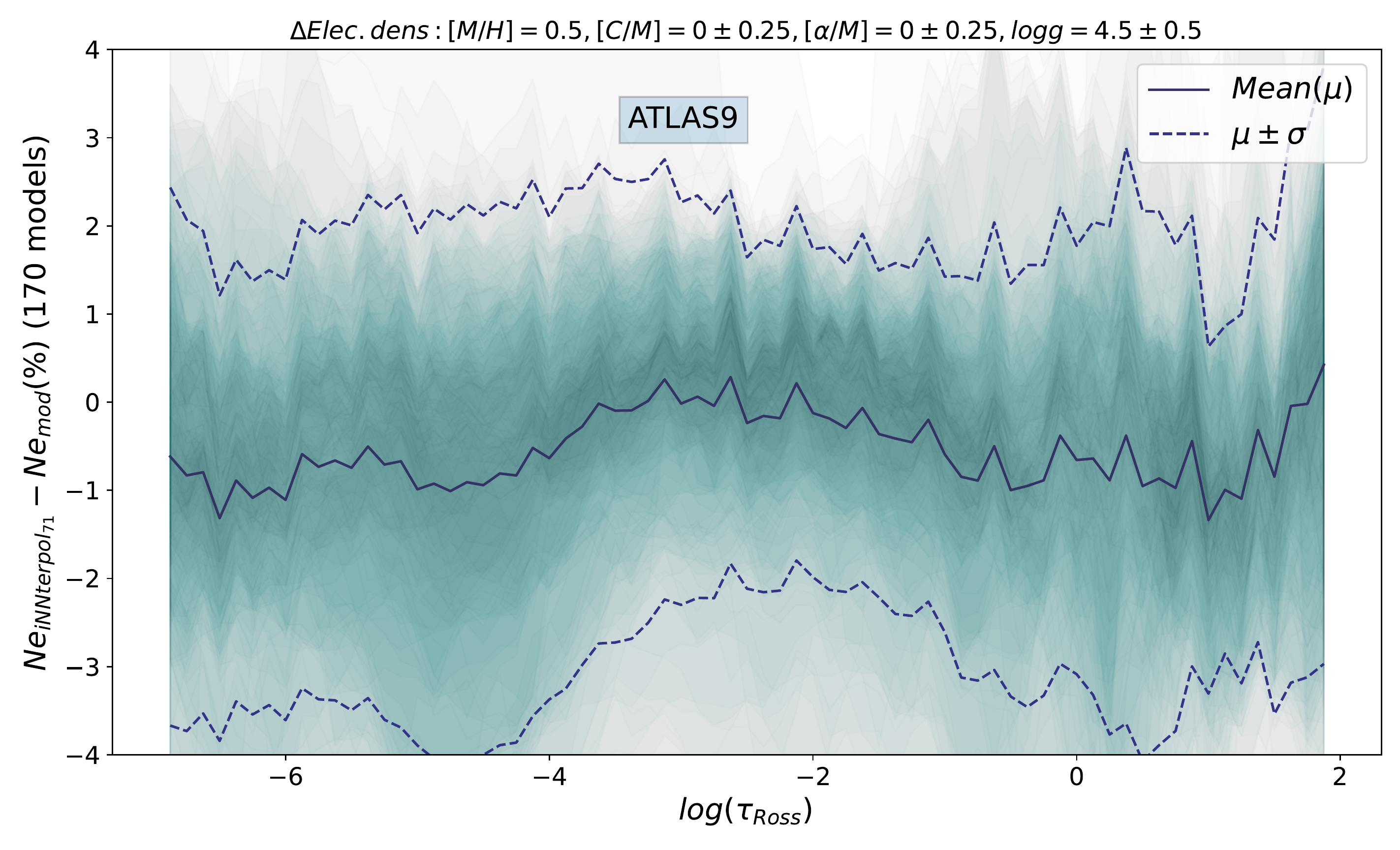}
    \caption{iNNterpol with a bottleneck of 71 (CAE71-NN) differences for electronic density as a percentage between the predicted models and the actual ones for points in the grid never seen for the NN for ATLAS9 model data.}
    \label{fig:DDens_chkCAE_16_71_M0_mp05_lgg45_scale}
\end{figure}

\begin{figure}
        \includegraphics[width=\columnwidth]{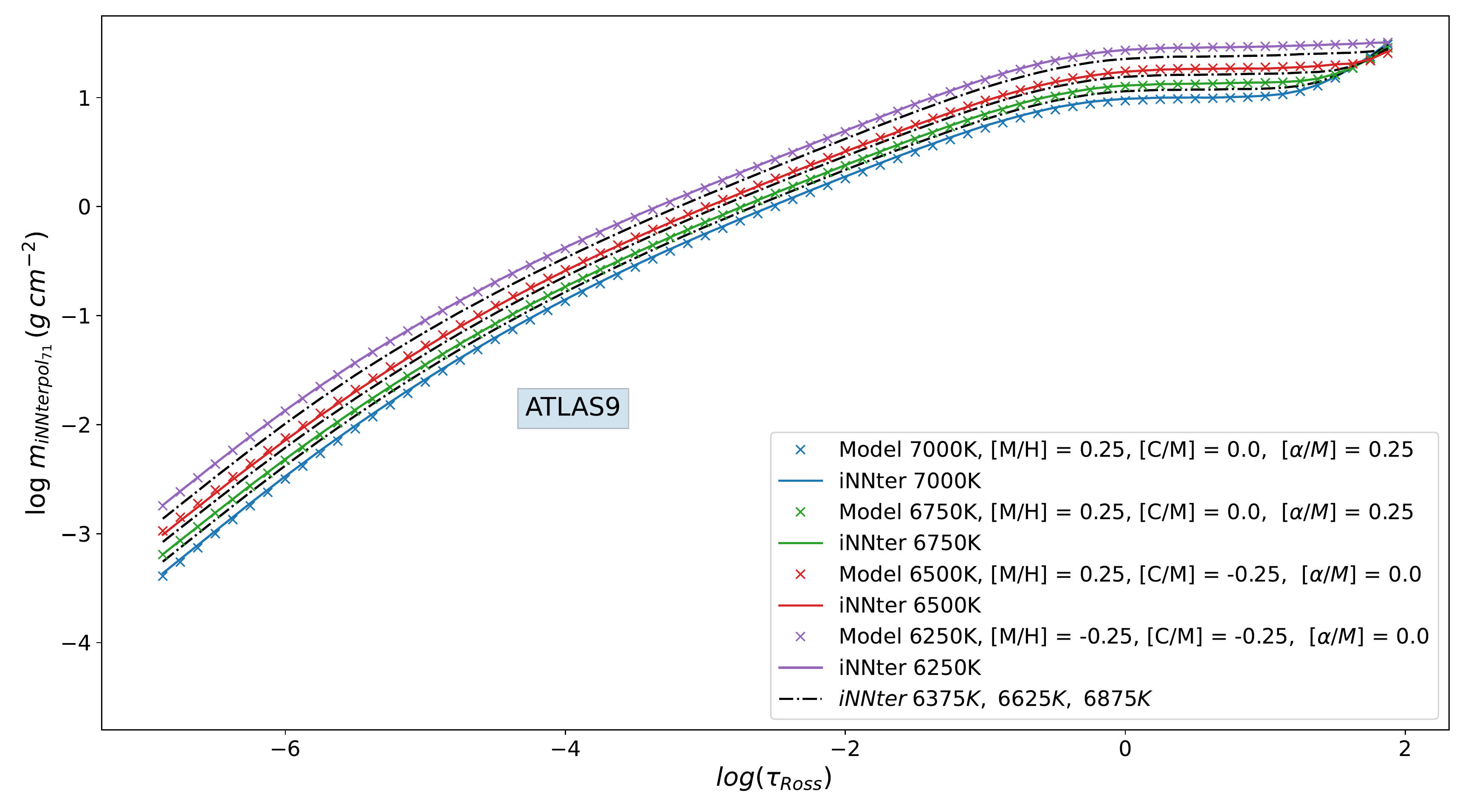}
    \caption{Same as Fig.~\ref{fig:Temp_CAE_16_71_Teff7000_6250_lgg20}, but for the 
    mass column.}
    \label{fig:Mass_CAE_16_71_Teff7000_6250_lgg20}
\end{figure}

\begin{figure}
        \includegraphics[width=\columnwidth]{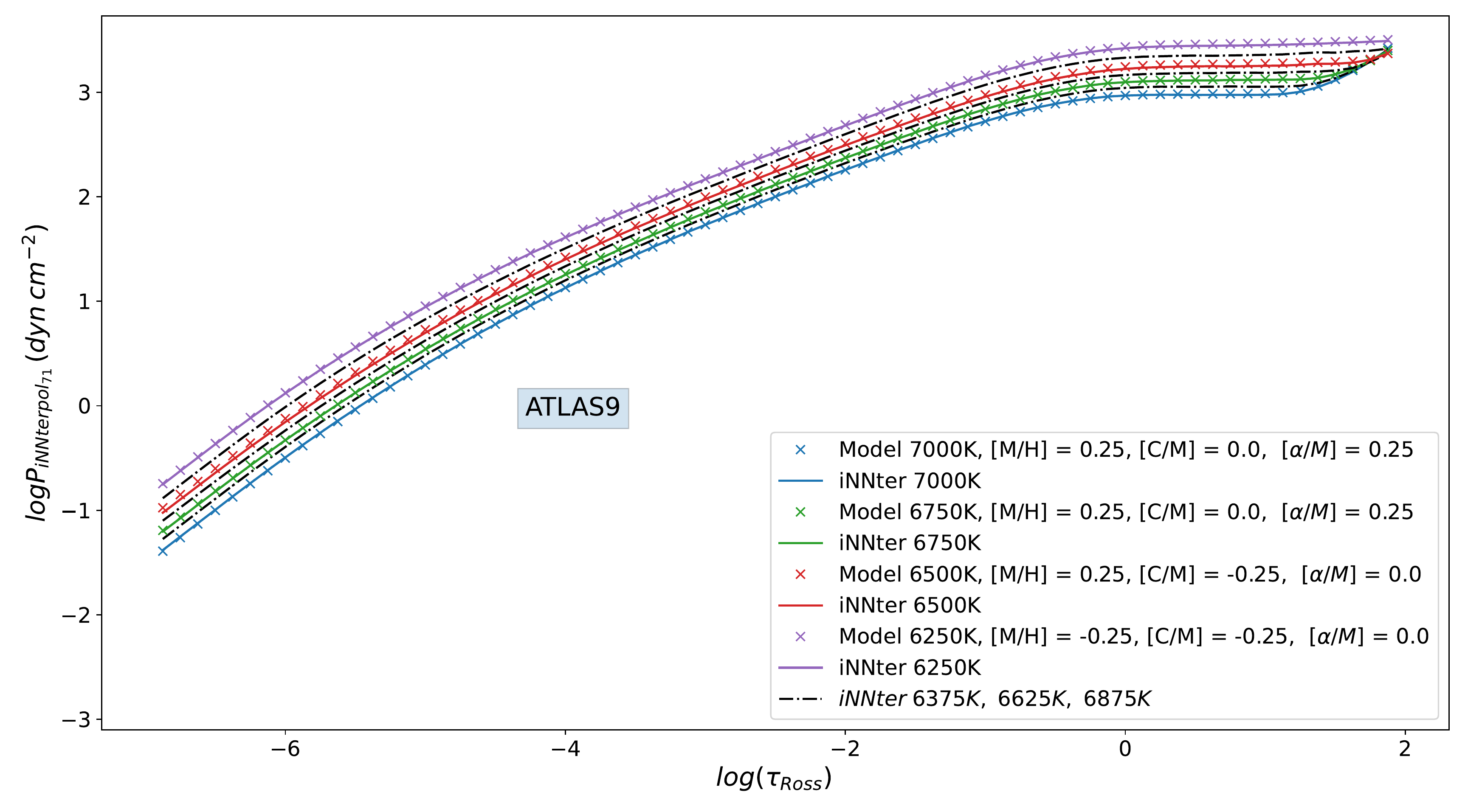}
    \caption{Same as Fig.~\ref{fig:Temp_CAE_16_71_Teff7000_6250_lgg20}, but for the gas pressure.}
    \label{fig:Pres_CAE_16_71_Teff7000_6250_lgg20}
\end{figure}

\begin{figure}
        \includegraphics[width=\columnwidth]{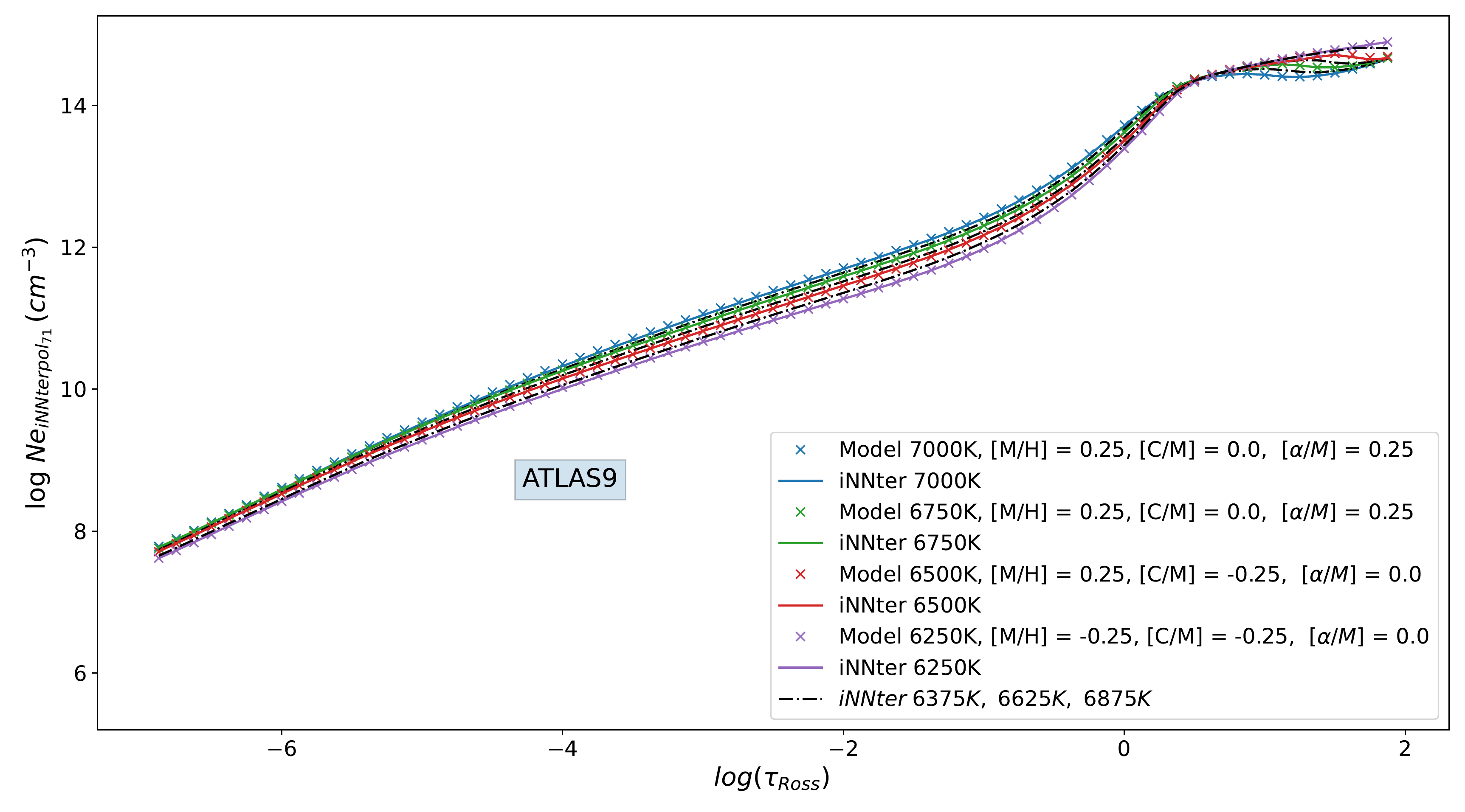}
    \caption{Same as Fig.~\ref{fig:Temp_CAE_16_71_Teff7000_6250_lgg20}, but for the electronic number density.}
    \label{fig:Dens_CAE_16_71_Teff7000_6250_lgg20}
\end{figure}

\begin{figure}
        \includegraphics[width=\columnwidth]{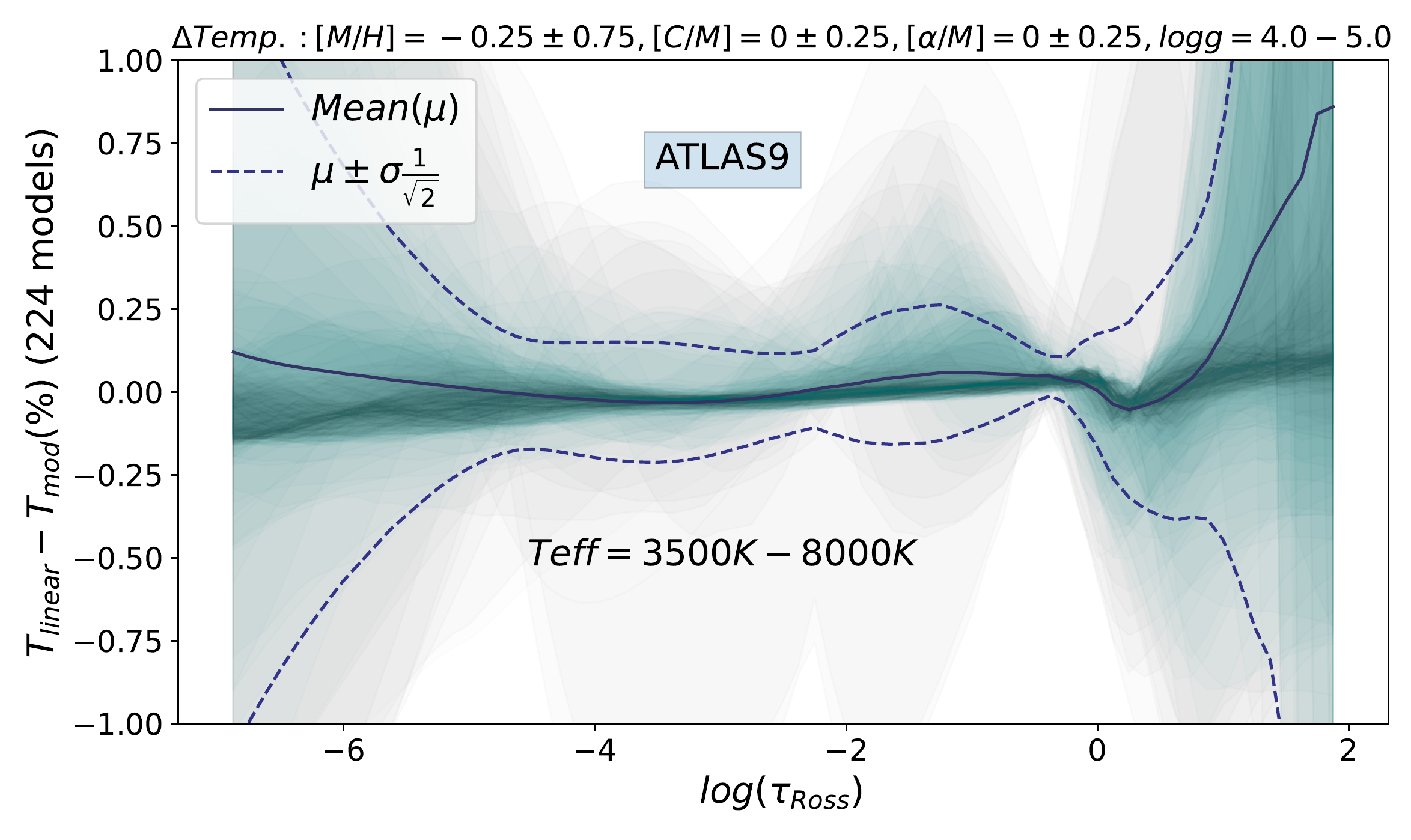}
    \caption{Errors of linearly interpolating for ATLAS9 cool dwarf stars.}
    \label{fig:DTemp_Linear_ATLAS9_mn10_mp05_lgg40_lgg50_Teff35k_Teff8k_cold_1pct}
\end{figure}

\begin{figure}
        \includegraphics[width=\columnwidth]{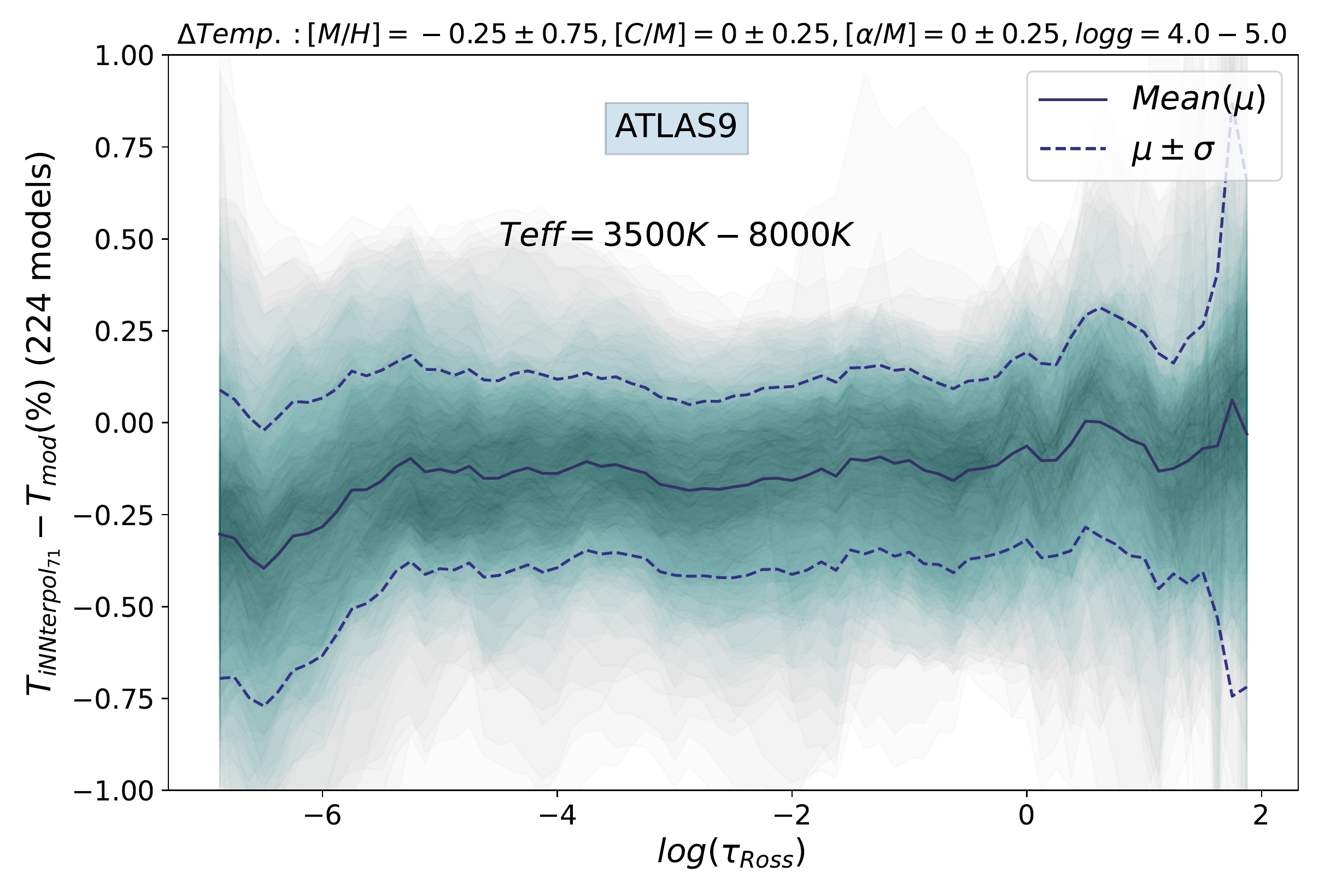}
    \caption{Errors of iNNterpol (CAE71-NN) for the same grid models as in Fig.~\ref{fig:DTemp_Linear_ATLAS9_mn10_mp05_lgg40_lgg50_Teff35k_Teff8k_cold_1pct} for direct comparison.}
    \label{fig:DTemp_chkCAE_16_71_ATLAS_mn10_mp05_logg4_logg5_Teff35k_Teff8k_cold_1pct}
\end{figure}

\begin{figure}
        \includegraphics[width=\columnwidth]{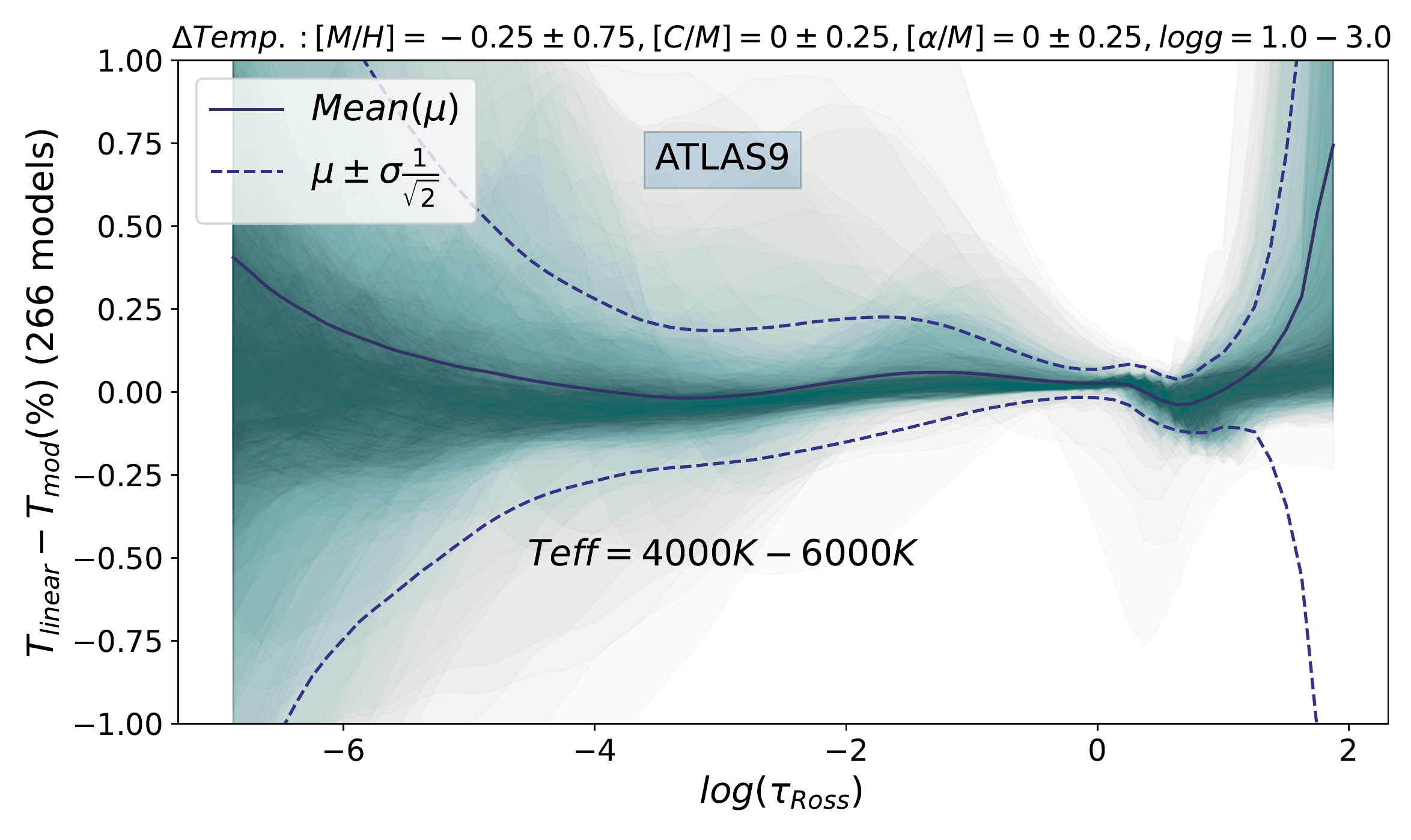}
    \caption{Errors of linearly interpolating for ATLAS9 giant stars.}
    \label{fig:DTemp_Linear_ATLAS9_mn10_mp05_lgg10_lgg30_Teff4k_Teff6k_giants_1pct}
\end{figure}

\begin{figure}
        \includegraphics[width=\columnwidth]{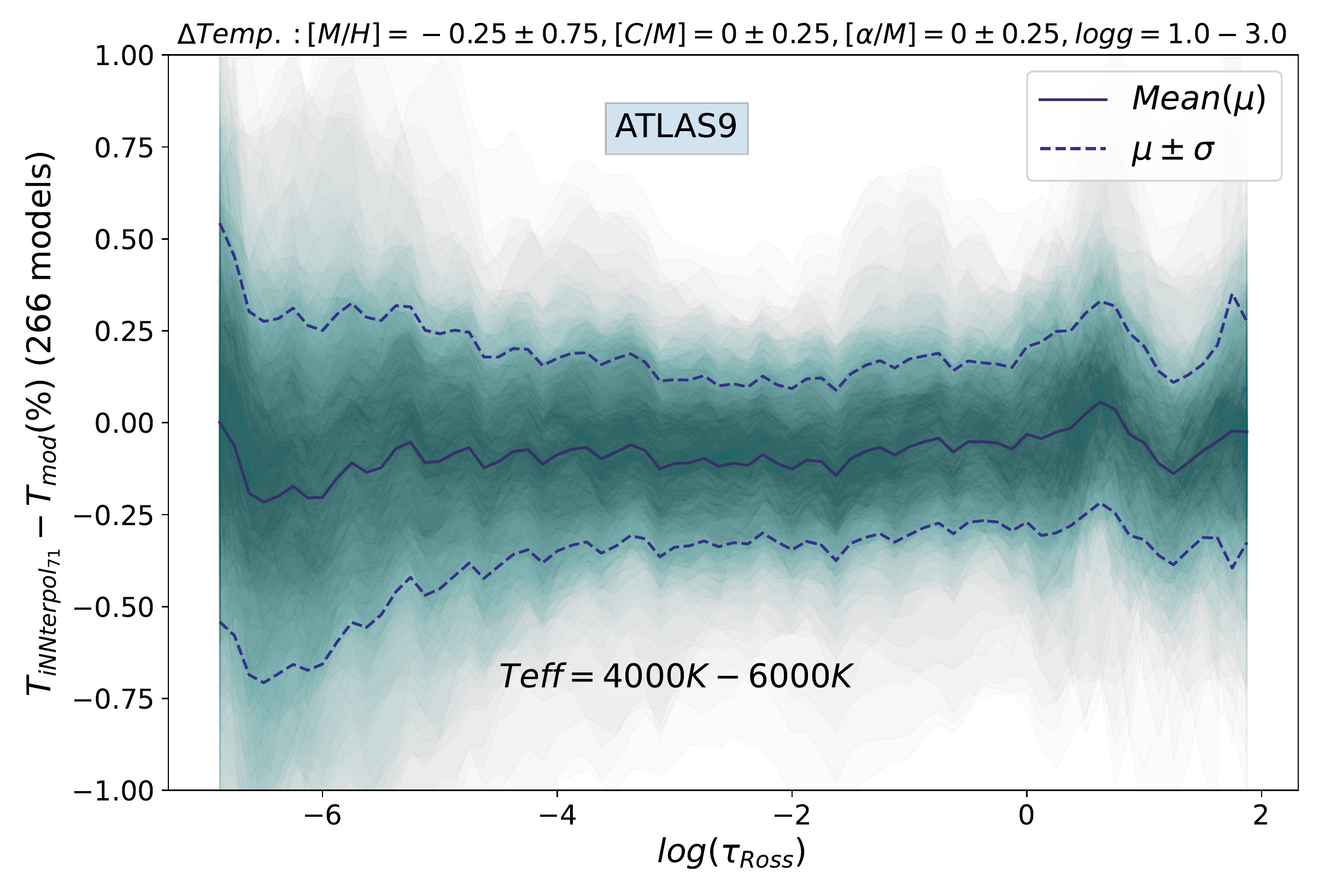}
    \caption{Errors of iNNterpol (CAE71-NN) for the same grid models as in Fig.~\ref{fig:DTemp_Linear_ATLAS9_mn10_mp05_lgg10_lgg30_Teff4k_Teff6k_giants_1pct} for direct comparison.}
    \label{fig:DTemp_chkCAE_16_71_ATLAS_mn10_mp05_logg10_logg30_Teff4k_Teff6k_giants_1pct}
\end{figure}

\end{appendix}

\end{document}